\documentclass[aps,prb,amsmath,amsfonts,amssymb,twocolumn, superscriptaddress,footinbib,showpacs, 10pt]{revtex4}
\usepackage{color}
\usepackage{graphicx}
\usepackage{epsfig}
\usepackage{epstopdf}
\usepackage{soul}

\newcommand*{\be}{\begin{equation}}
\newcommand*{\ee}{\end{equation}}
\newcommand*{\bea}{\begin{eqnarray}}
\newcommand*{\eea}{\end{eqnarray}}

\newcommand{\cecoin}{CeCoIn$_5$}
\newcommand{\cerhin}{CeRhIn$_5$}

\begin{document}

\title{Two-channel point-contact tunneling theory of superconductors}

\author{Mikael Fogelstr\"om} 
\affiliation{Department of Microtechnoloy and Nanoscience, Chalmers, S-412 96 G\"oteborg, Sweden}
\author{Matthias J. Graf}
\affiliation{Theoretical Division, Los Alamos National Laboratory, Los Alamos, New Mexico 87545, USA}

\author{V. A. Sidorov}
\affiliation{Vereschagin Institute of High Pressure Physics, RAS, 142190 Troitsk, Russia}
\author{Xin Lu}
\affiliation{Center for Correlated Matter, Zhejiang University, Hangzhou 310058, China}
\author{E. D. Bauer}
\affiliation{Materials Physics and Applications Division, Los Alamos National Laboratory, Los Alamos, New Mexico 87545, USA}
\author{J. D. Thompson}
\affiliation{Materials Physics and Applications Division, Los Alamos National Laboratory, Los Alamos, New Mexico 87545, USA}

\date{\today}

\begin{abstract}
We introduce a two-channel tunneling model to generalize the widely used BTK theory of point-contact conductance 
between a normal metal contact and superconductor. 
Tunneling  of electrons  can occur via localized surface states or directly, resulting in a Fano resonance in the differential conductance $G=dI/dV$.
We present an analysis of $G$ within the two-channel model when applied to soft point-contacts between  normal metallic silver particles and prototypical heavy-fermion superconductors \cecoin\ and \cerhin\ at high pressures.
In the normal state the Fano line shape of the measured $G$ is  well described by a 
model with two tunneling channels and a large temperature-independent background conductance.
In the superconducting state a strongly suppressed Andreev reflection signal is explained by the presence of the background conductance.
We report Andreev signal in \cecoin\ consistent with standard $d_{x^2-y^2}$-wave pairing, 
assuming an equal mixture of tunneling into [100] and [110] crystallographic interfaces. Whereas in \cerhin\ at 1.8 and 2.0 GPa the signal is described by a
$d_{x^2-y^2}$-wave gap with reduced nodal region, i.e., increased slope of the gap opening on the Fermi surface.
A possibility is that the shape of the high-pressure Andreev signal is affected by 
the proximity of a 
line of quantum critical points that extends from 1.75 to 2.3 GPa,
which is not accounted for in our description of the heavy-fermion superconductor.
\end{abstract}

%

\maketitle
\section{Introduction}

There has been considerable work exploring the complex phase diagram of  the heavy fermion \cerhin.
\cite{Hegger2000, Mito2001, Fisher2002, Higemoto2002, Mito2003, Llobet2004, Shishido2005, Park2006, Park2008a, Park2008b, Park2008c, Park2011} 
As function of pressure the antiferromagnetism (AFM) is suppressed toward a quantum-critical state and superconductivity (SC)
appears. Early on it has been speculated that a quantum critical point
(QCP) is at the heart of electron pairing mediated by strong magnetic fluctuations.\cite{Mathur1998,Si2001,Senthil2004,Coleman2005} 
This scenario could explain that above a critical pressure of about
1.75 GPa in \cerhin\ specific heat and NQR measurements find
a pure superconducting phase,
while below the AFM and SC coexist.\cite{Mito2003,Park2006}

The observation of power-law temperature dependence of 
the spin-lattice relaxation rate and thermodynamic
properties down to 0.3 K in \cerhin\ at high pressures
has been taken as evidence for nodal quasiparticle states and firmly established the similarity
between \cerhin\ and \cecoin. \cite{Mito2001, Fisher2002, Higemoto2002}
In fact, recent scanning tunneling spectroscopy measurements provide direct evidence for nodal states in agreement with a $d_{x^2-y^2}$-wave superconducting order parameter in \cecoin.\cite{Zhou2013, Allan2013}
The additional observation of field-angle-dependent fourfold oscillations in the specific heat of \cerhin,\cite{Park2008c} similar to
\cecoin,\cite{Izawa2001, Aoki2004, An2010} has been interpreted in favor of a $d_{x^2-y^2}$-wave gap.  
While the observation of fourfold oscillations and sign reversal of the oscillation amplitude in the specific heat
are consistent with the $d_{x^2-y^2}$-wave gap scenario for \cecoin\ and \cerhin\ within a two-band model of superconductivity, 
the unique identification of 
$d_{x^2-y^2}$-wave symmetry at temperatures above roughly one-fifth of the superconducting transition temperature is complicated
due to the competing effects of Fermi surface anisotropy.\cite{Das2013}

Unambiguous determination of the symmetry of an unconventional superconducting order parameter is difficult. However, one can obtain
crucial information about the superconducting gap from point-contact spectroscopy (PCS). Under favorable conditions the line shape of an Andreev 
reflection signal or zero-bias conductance peak may allow further differentiation between possible pairing symmetries.
In many heavy-fermion materials this is further complicated due to asymmetric Fano-like line shapes in the normal-state conductance
on top of a strongly suppressed Andreev reflection signal. \cite{Nowack1987,DeWilde1994,Naidyuk1996,naidyuk1998}  
This has been especially true for the heavy-fermion superconductor (HFS) \cecoin. \cite{parkwk2005,goll2005,goll2006,parkwk2008a,parkwk2008b,parkwk09}

Phase coherence and Andreev reflection are considered hallmarks of superconductivity, because they require the existence of a Cooper pair condensate.
Andreev reflection occurs only when a quasiparticle retro-reflects off a normal-superconducting (NS) interface as a quasihole, while 
momentum and charge are conserved and carried across the interface by the Cooper pair. Thus detection of an Andreev signal is a property unique to superconductors.
The Blonder-Tinkham-Klapwijk (BTK) theory describes the $dI/dV$ curve in conventional NS junctions by
invoking a dimensionless barrier strength parameter, which depends on the barrier potential and the mismatch ratio of 
Fermi velocities.\cite{btk} 
However, for HFS this formula predicts that normal-HFS (NHFS) junctions are in the tunneling limit (low transparency),
where Andreev reflection cannot occur, contrary to experimental observations. Nevertheless the BTK formula has been widely applied to this unphysical regime
due to the lack of alternative expressions.
Numerous attempts have been proposed to correct the BTK formula, but with limited success.\cite{Deutscher1994,Gloos1996,Anders1997}
In Ref.~\onlinecite{Fogelstrom2010} we proposed a multichannel tunneling model for PCS to circumvent the inherent shortcomings of the BTK theory. With this multichannel model a consistent description of normal- and superconducting state PCS data  was possible in the HFS \cecoin.

Recently, we developed a technique to measure the differential tunneling conductance $G=dI/dV$
of a superconductor at high pressure by  PCS. Since the contact is formed {\it gently}
with a coating of silver particles, this technique is also called in the literature soft point-contact spectroscopy (SPCS) to distinguish it from conventional
metal tip PCS (see e.g. the review Ref.~\onlinecite{Daghero2010} and references within).
This novel SPCS technique allows the study of
the pressure-dependent electronic properties of normal metals and superconductors as characterized by the $dI/dV$ curve.
In superconductors it provides crucial information about the opening of the excitation gap of Cooper pairs,
where the PCS technique is often the first measurement to determine the magnitude of the superconducting gap. 

In this paper, we use SPCS to address the question of the pairing symmetry in \cerhin. We ask if the 
pairing symmetry is the same across the pressure phase diagram
and if it is the same $d$-wave symmetry as in the sister compound \cecoin. 
A direct measurement of the superconducting gap structure near the coexistence region and deeper into the superconducting dome might provide the necessary answers toward the importance of the QCP around $P=1.75$ GPa.

The paper is organized as follows. 
In Sec.~II we introduce the point-contact tunneling theory and two-channel tunneling formulas for the normal and superconducting state.
In Sec.~III we present theoretical results of three typical tunneling regimes relevant to point contacts
and analyze the SPCS data of \cecoin\ and \cerhin. The spectra are discussed and interpreted in light of our two-channel model.
We conclude our results in Sec.~IV.

\section{Point-Contact Tunneling Theory}

We follow closely our earlier work in Ref.~\onlinecite{Fogelstrom2010}, except with the simplification that here we consider only a single band of itinerant electrons in the HFS,
even though de Haas-van Alphen measurements and electronic structure calculations reveal several Fermi surface sheets.\cite{Settai2001, Hall2001, Elgazzar2004}
The itinerant electrons are characterized by Fermi surface parameters and the localized surface states have a single energy level $E_0$. We choose the Fermi level $E_f=0$.
One may justify the existence of localized surface states through the mechanical process of contact making between any metallic object and the system of interest, since such a contact deforms the surface and thus breaks its translational symmetry, 
or through the presence of impurities.
Finally, we allow the itinerant electrons to condense into a superconducting ground state.

The standard tunneling Hamiltonian of the point contact with a conductor, and in particular a heavy-fermion system ${\cal H}_{\rm HF}$, is the combination of the tip (contact) and the transfer (tunneling) processes between them:
${\cal H} = {\cal H}_{\rm HF} + {\cal H}_{\rm tip} + {\cal H}_{\rm T}$.
The tip is given by normal conduction electrons in the contact
and the tunneling Hamiltonian ${\cal H}_{\rm T}$ describes all possible transfers.
In addition to the standard overlap integral between the conduction band in 
the point contact and the itinerant band of the conductor, $t$,   there is finite overlap
from the point contact to localized states, $t_{loc}$. 
Weak interaction between localized surface electrons and itinerant electrons is accounted for through the scattering (hybridization) term $v$.
The model is illustrated in Fig.~\ref{fig:cartoon_hamiltonian}(a) and a detailed description is given in the Appendix.
Finally, in order to attain a Fano resonance in the conductance one needs to include quantum interference between different tunneling 
paths, e.g., interference between an electron from the metal tip to the localized state and on to the itinerant electrons in the conductor versus a direct pathway between contact and itinerant electrons.\cite{fano1961}

\begin{figure}[bth]
\includegraphics[width=0.95\columnwidth,angle=0]{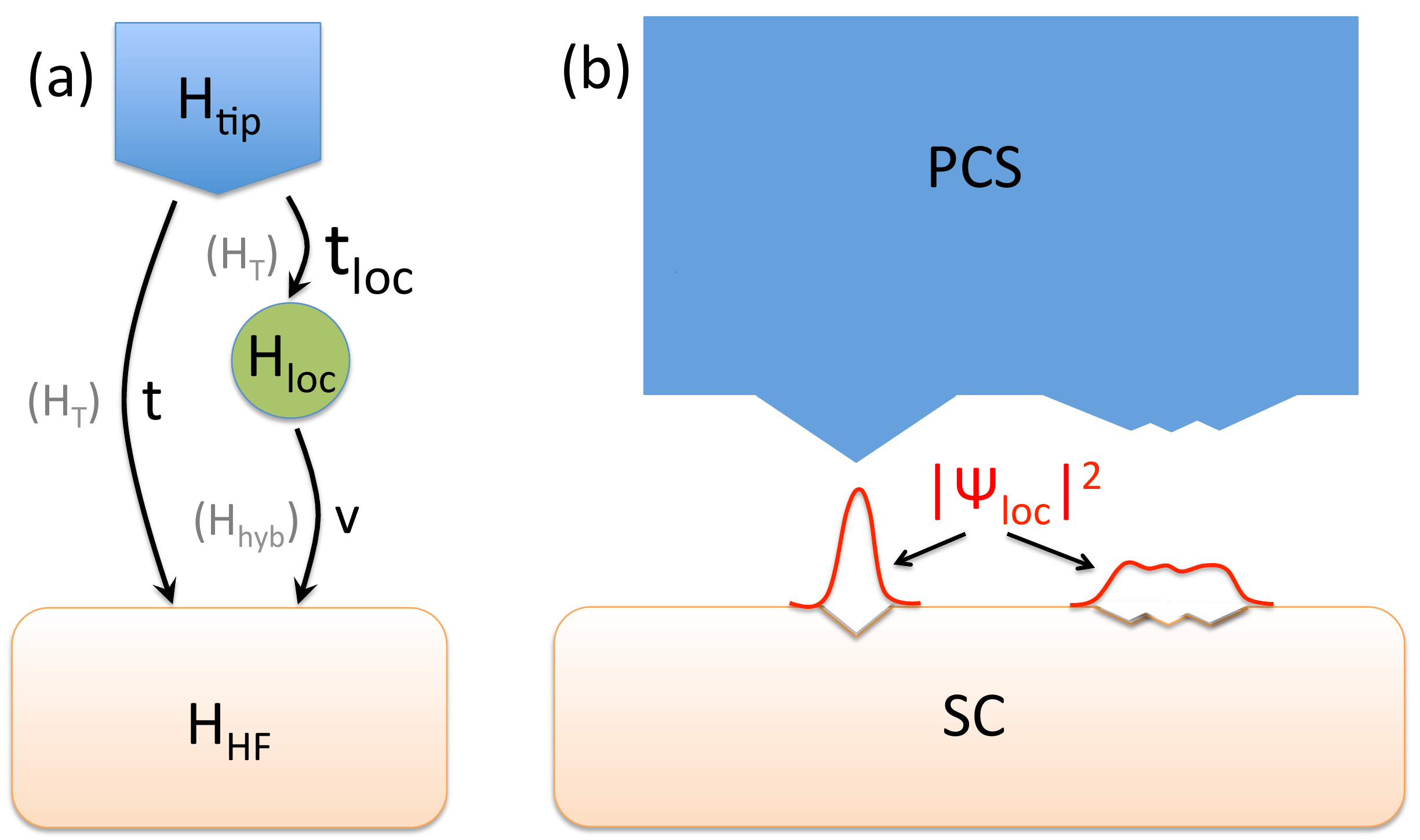}
\caption{(color online) (a) Cartoon of all contributions to the standard tunneling Hamiltonian of a point contact with overlap integrals ($t, t_{loc}, v$).
(b) Schematics of surface defects generated by the PCS contact with atomically rough metallic tip and corresponding wave function amplitudes $|\Psi_{loc}|^2$ of ``sharp'' and ``broad''  localized defect states. The tunneling conductance will be governed by the stronger overlap between the wave functions of the tip
and   ``sharp'' defect states compared to ``broad'' surface states. 
Here we suggest that the radial wave function of  ``sharp''  localized states extends further into the open space, normal to the surface of the (heavy-fermion) superconductor.
}
\label{fig:cartoon_hamiltonian}
\end{figure}

\subsection{Normal State Point-Contact Tunneling}

In Ref.~\onlinecite{Fogelstrom2010} we derived an analytic expression for the multiband
tunneling current using the standard Green's function method.\cite{HaugJauho, Cuevas}
We showed that in Keldysh notation the tunneling current of the point contact can be written in a
compact notation (see Appendix for further details):
\begin{eqnarray}
I(V) &=& \frac{e}{\hbar}{\rm Tr}\, \hat\tau_3 
\bigg\lbrack 
	\check t_{loc}\circ \check G_{loc,c}-\check t_{loc}^*\circ \check G_{c,loc}
\nonumber\\ && \quad
	+ \check t \circ\check G_{h,c}-\check t^*\circ \check G_{c,h} 
\bigg\rbrack^K.
\label{eq:tunnelingcurrent}
\end{eqnarray}
Here the trace (Tr) is a short-hand notation for summation over momentum $k$ and spin $\sigma$. 
The $\circ$-product indicates a folding over common arguments, e.g., 
$\check t_{loc}\circ \check G_{loc,c}(k,k^\prime)=\sum_{k^{\prime}} t_{loc,kk^\prime} \check G_{loc,c}(k^\prime,k^{\prime\prime})$,
and $\lbrack\,\,\rbrack^K$ denotes the Keldysh component of the matrix Green's function.
Notation for matrices is: a ``hat" ($\hat x$) denotes a Nambu matrix, while a ``check" ($\check x$) represents a Keldysh matrix. 
In equation (\ref{eq:tunnelingcurrent}), $\check G_{i, j}$ are Green's function components of the
full matrix in reservoir space ($c$=point contact, $h$= heavy conduction band of the HFS, and
``$loc$" is the localized surface state).
The components of $\check G$  straddle the interface
(e.g. $ \check G_{c, h}, \check G_{c, loc}$,  etc.) 
and were determined previously in the presence of a voltage bias across the interface.\cite{Fogelstrom2010}
In the remainder of this work, we will use these previously derived solutions.

In addition to proposing localized surface states in Ref.~\onlinecite{Fogelstrom2010}, we hypothesized that the differential conductance of a PCS contact is made of many quantum channels with sharp and smeared out localized states,
as illustrated in Fig.~\ref{fig:cartoon_hamiltonian}(b).
The sharp localized states superpose to give the Fano line shape, while the broad localized states contribute to the 
weakly voltage-dependent background conductance.
This approximation led us to the Fano expression within the tunneling model. 
In this work we build on the results of Ref.~\onlinecite{Fogelstrom2010} and start with the general expression for the differential conductance, 
$G \equiv \frac{dI}{dV}$,
\begin{eqnarray}
G  &=& \frac{ C_0 }{ {\cal D} }
\, 
\int_{n_S \cdot p < 0} dp
\int^\infty_{-\infty} \frac {d \varepsilon}{4 T}
{\rm sech}\bigg\lbrack\frac{\varepsilon-eV}{2T}\bigg\rbrack^{2}
K(p, \varepsilon) +
\nonumber\\ && 
G_0(V). 
\label{eq:G}
\end{eqnarray}
The momentum integration is performed over the half space $n_S \cdot p < 0$ with the superconductor's surface normal $n_S$ and Fermi surface momentum $p$.
The parameter $C_0$ and background function $G_0(V)$ are determined by the large-voltage scale of the conductance of the contact.
The factor $C_0$ is proportional to the transparency ${\cal D}$ and the quantum conductance of the single tunneling channel
($G_Q \equiv 2e^2/h=0.07748$ 1/k$\Omega$)
times the total number of quantum channels of the point contact.
In SPCS experiments potentially more than $10^4$ ($\sim G_0(0)/G_Q$) conductance channels contribute, though far fewer,
namely on the order of $10^2$ ($\sim C_0/G_Q$), dominate the low-voltage tunneling $dI/dV$ characteristics.
In principle, these unknowns could be obtained from a microscopic theory of the distribution of sharp and broad localized tunneling channels.
However, for simplicity, we treat them as fit parameters and fit functions.
Other contributions to the background conductance $G_0(V)$ might come from additional conduction bands, which are weakly coupled to the metal tip, or from a strongly energy-dependent density of states near the Fermi level.
For simplicity, we will not consider these possibilities in order to keep the problem tractable.

Following the notation of Ref.~\onlinecite{Fogelstrom2010} the  normal-state conductance kernel 
is momentum independent and
can be written as
\begin{eqnarray}
K(p, \varepsilon) = 
{\cal D}
\frac{(q_F \Gamma +\varepsilon-{E})^2}{\Gamma^2+(\varepsilon-{E})^2 } ,
\label{eq:kernel_NS}
\end{eqnarray}
where ${E}$ is now the tunneling-renormalized 
value of the localized energy level $E_0$, $\Gamma$ is the half-width of the resonance,
and $q_F$ is the Fano quantum interference parameter that controls the Fano resonance line shape. 
These phenomenological parameters
can be extracted from the PCS experiment and determine our microscopic model parameters, 
\begin{eqnarray}
\label{eq:parameters_begin}
{\cal D}&=&\frac{4 t^2}{(1+ t^2)^2} ,
\\
E &=&{E}_0-\frac{2 t_{loc} v t}{1+t^2} ,
\\
\Gamma&=&\frac{t_{loc}^2+v^2}{1+t^2} ,
\\
q_F&=& \frac{ t_{loc} v  t}{ t^2_{loc} + v^2 }\frac{1-t^2}{t^2} .
\label{eq:parameters_end}
\end{eqnarray}

In Ref.~\onlinecite{Fogelstrom2010}, we compactified our notation and moved the density-of-states factors
into the tunneling elements.\cite{qFsign} Thus the new and renormalized tunneling element $t$ is dimensionless, while $t_{loc}$
and $v$ have dimension $\sqrt{\rm energy}$.
This treatment is equivalent to saying that the density of states at the Fermi level is flat in both metal tip and heavy-fermion conductor. While this approximation is typically justified for normal metals, it is not obvious why it should hold for heavy fermions with narrow $f$ bands.
This short-coming can be overcome by incorporating the energy dependence of the density of states in the HFS following the work of Ref.~\onlinecite{HaugJauho}.
However,  we would have to pay the price of losing the transparency and simplicity of the normal-state conductance kernel in Eq.~(\ref{eq:kernel_NS}).
Note that in the limit of $t\to 0$ the Fano kernel in Eq.~(\ref{eq:kernel_NS}) reduces to that of a Lorentzian, which is maximum for resonant ($E_0=0$) tunneling.

In our analysis of the PCS conductance we use the  formula in Eq.~(\ref{eq:G}) with the normal-state kernel in (\ref{eq:kernel_NS}) to extract the temperature-dependent phenomenological model parameters and the temperature-independent background function $G_0(V)$ from a set of PCS measurements at different temperatures. 
Once the parameters in Eqns.~(\ref{eq:parameters_begin})-(\ref{eq:parameters_end}) have been fit, we map them onto the corresponding microscopic model parameters. Since ${\cal D}$ is only proportional to $C_0$ times an unknown contact area or number of quantum tunneling channels, the mapping between $(\Gamma, q_F, E)$ and $(t, t_{loc}, v, E_0)$ at any given temperature is not unique until the Andreev reflection signal is measured in the superconducting state.

Another interesting aspect of  the normal-state conductance kernel in Eq.~(\ref{eq:G}) is its invariance under the exchange of the tunneling parameters $t_{loc}$ and $v$. Since the Eqns.~(\ref{eq:parameters_begin})-(\ref{eq:parameters_end}) are symmetric under $t_{loc} \leftrightarrow v$,  it is impossible to distinguish an adatom or impurity on the contact tip from one on the sample surface.

\subsection{Superconducting State Point-Contact Tunneling}

Entering the superconducting state the conductance is significantly changed and highly nonlinear in voltage,
as well as depends on momentum $p$.
However, it is still possible to generalize the normal-state expression in Eq.~(\ref{eq:kernel_NS}) to include superconductivity. After tedious but straightforward regrouping of terms in the expression of the tunneling current, the superconducting conductance kernel can be written in compact form
\begin{eqnarray}
K(p,\varepsilon)
&=&(1+|{\cal R}_p|)(q_F\Gamma_e+\varepsilon- E_{e})^2 
 \nonumber\\
&&\times\frac{ D_+- D_-(1-{\cal D})\, |{\cal R}_p| }{| A_+ A_-+(1-{\cal D}) {\cal R}_p B_+B_-|^2}
\label{eq:kernel_SC}
\end{eqnarray}
where we introduced the coefficients
\bea
A_{\pm}&=&\varepsilon \pm 
\left(
E_0-\frac{2t_{loc}v t}{1+t^2}
\right)
+i \frac{t_{loc}^2+v^2}{1+t^2}\nonumber\\&=&\varepsilon\pm E_{e}+i\Gamma_e ,\\
B_{\pm}&=&\varepsilon \pm 
\left(
E_0+\frac{2t_{loc}v t}{1-t^2}
\right)
+i \frac{t_{loc}^2-v^2}{1-t^2}\nonumber\\&=&\varepsilon\pm E_{h}+i\Gamma_h ,\\
D_{\pm}&=&\bigg(\frac{t_{loc}^2\pm v^2}{1\pm t^2}\bigg)^2+
\left(
\varepsilon+E_0\mp\frac{2t_{loc}v t}{1\pm t^2}
\right)^2 
\nonumber\\
&=&\Gamma^2_{e/h}+(\varepsilon+ E_{e/h})^2 ,
\eea
and the momentum-dependent Andreev reflection probability
\be
{\cal R}_p =\gamma(p_{in},\varepsilon) \tilde \gamma(p_{out},\varepsilon),
\ee
which carries information about the superconducting condensate via the coherence factors
\bea
&&\gamma(p,\epsilon)=\frac{-\Delta(p)}{\varepsilon+i\sqrt{|\Delta(p)|^2-(\varepsilon^{R})^2}} ,\\
&&\tilde\gamma(p,\epsilon)=\frac{\Delta^*(p)}{\varepsilon+i\sqrt{|\Delta(p)|^2-(\varepsilon)^2}} .
\eea

Here $p$ is a momentum on the Fermi surface and $\Delta(p)$ is the gap function.
In the usual convention, $p_{in}$ is the momentum of a quasiparticle 
moving into the NS interface coming from the superconductor and $p_{out}$ describes the opposite process.
Since we assume specular reflection in all our calculations, the ``{\it in}'' and ``{\it out}'' momenta are related by
$p_{out} = p_{in} - 2  n_S \cdot p_{in}$,  with the superconductor's surface normal $n_S$.
At first sight it appears that the conductance kernel $K(p, \varepsilon)$ diverges at $t\to1$ because of coefficients $B_\pm$ and $D_\pm$.
However, this is not the case because the factor $(1-{\cal D})=[(1-t^2)/(1+t^2)]^2$ regularizes the perceived divergence.
Note that for $\Delta(p) = 0$ the superconducting kernel in Eq.~(\ref{eq:kernel_SC})
reduces to the normal-state kernel in Eq.~(\ref{eq:kernel_NS}).

Unlike in the normal state the conductance kernel in the superconducting state in Eq.~(\ref{eq:kernel_SC}) breaks the symmetry $t_{loc} \leftrightarrow v$, because of the coefficients $B_\pm$ or more precisely because of the damping term $\Gamma_h$. Thus in principle it might be possible to distinguish between an adatom or impurity on the contact tip versus one on the sample surface by analyzing in detail the width of the line shape of the Andreev reflection signal.

\begin{figure}[bth]
\includegraphics[width=0.95\columnwidth,angle=0]{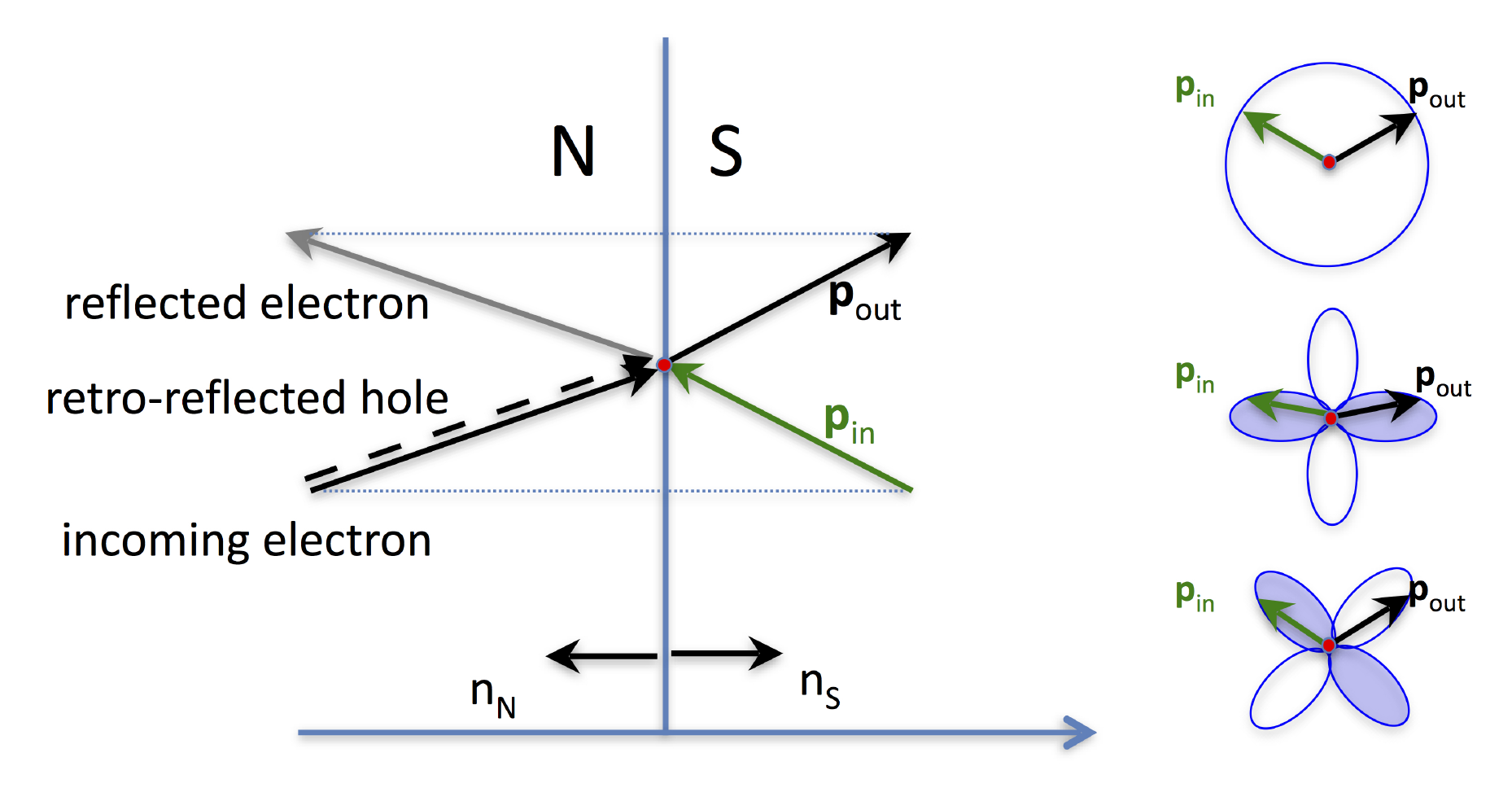}
\caption{(color online) Cartoon of the quasiparticle scattering processes at the NS interface between the metal tip and the superconductor (a).
Note that the velocity of the retro-reflected hole is antiparallel to that of the incoming electron, while both have the same momentum (direction of arrow).
The importance of the sign change of the gap function, connecting trajectories of the scattered quasiparticle in S with momenta $p_{in}$ and $p_{out}$,
 is shown for
(b) isotropic $s$, (c) $d_{xy}$, and (d) $d_{x^2-y^2}$ wave gap functions $\Delta(p)$ relative to the surface normal $n_S$.
}
\label{fig:cartoon_ns}
\end{figure}

The scattering processes at the NS interface are illustrated in Fig.~\ref{fig:cartoon_ns}.
When transmission is not perfect, an incoming electron from the metal tip with energy below the gap can be reflected or retro-reflected as a hole.
The latter process gives rise to excess conductance also known as Andreev reflection signal, while the former accounts for the suppression of the Andreev signal.
Only the momentum parallel to the interface is conserved (as indicated by the dotted lines) in the scattering process. In addition, the momenta of the in- and out-going 
scattering quasiparticle in the superconductor are related through the specular (perfect mirror) reflection condition. It is this condition that can connect positive and negative lobes of the superconducting gap function resulting in a sign change of the Andreev reflection probability ${\cal R}_p$.
For the given surface normal $n_S$ only the $d_{x^2-y^2}$ gap 
in Fig.~\ref{fig:cartoon_ns}(d) is maximal pairbreaking at the interface and leads to a sign change in ${\cal R}_p$ with significantly altered PCS conductance line shape
compared to $s$ and $d_{xy}$ gap functions.

We wish to emphasize that up to this point the formulation of our two-channel tunneling model is applicable to any point contact between a metal tip and superconductor in the presence of localized surface states and is not restricted to heavy-fermion superconductors. 
However, we believe that in compounds with $d$ and in particular with $f$ electrons the role of localized surface states may be more prominent than usual, because the radial wave functions of atoms with occupied $d$ and $f$ orbitals extend much further into space than, for example, when only lower shells are filled.
In subsections \ref{subsection_ns} and \ref{subsection_sc}, we will apply this general tunneling model to the Ce-115 family of heavy fermions.

\section{Results and Discussion}

The general conductance formula in Eq.~(\ref{eq:kernel_SC}) of the two-channel tunneling model includes the three widely studied regimes of (1) direct tunneling between the metal contact and superconductor, (2) tunneling through the localized state into the superconductor, and (3) interference tunneling through the localized state and directly into the superconductor. The three qualitatively different tunneling regimes will be discussed in more detail below.
In the Figs.~\ref{fig:fig_btk}, \ref{fig:fig_lorentz} and \ref{fig:fig_fano} we plot the conductance kernel (\ref{eq:kernel_SC}) when we discuss the generic $dI/dV$ at absolute zero temperature.

\subsection{BTK Line Shape Regime}

\begin{figure}[bth]
\includegraphics[width=1.0\columnwidth,angle=0]{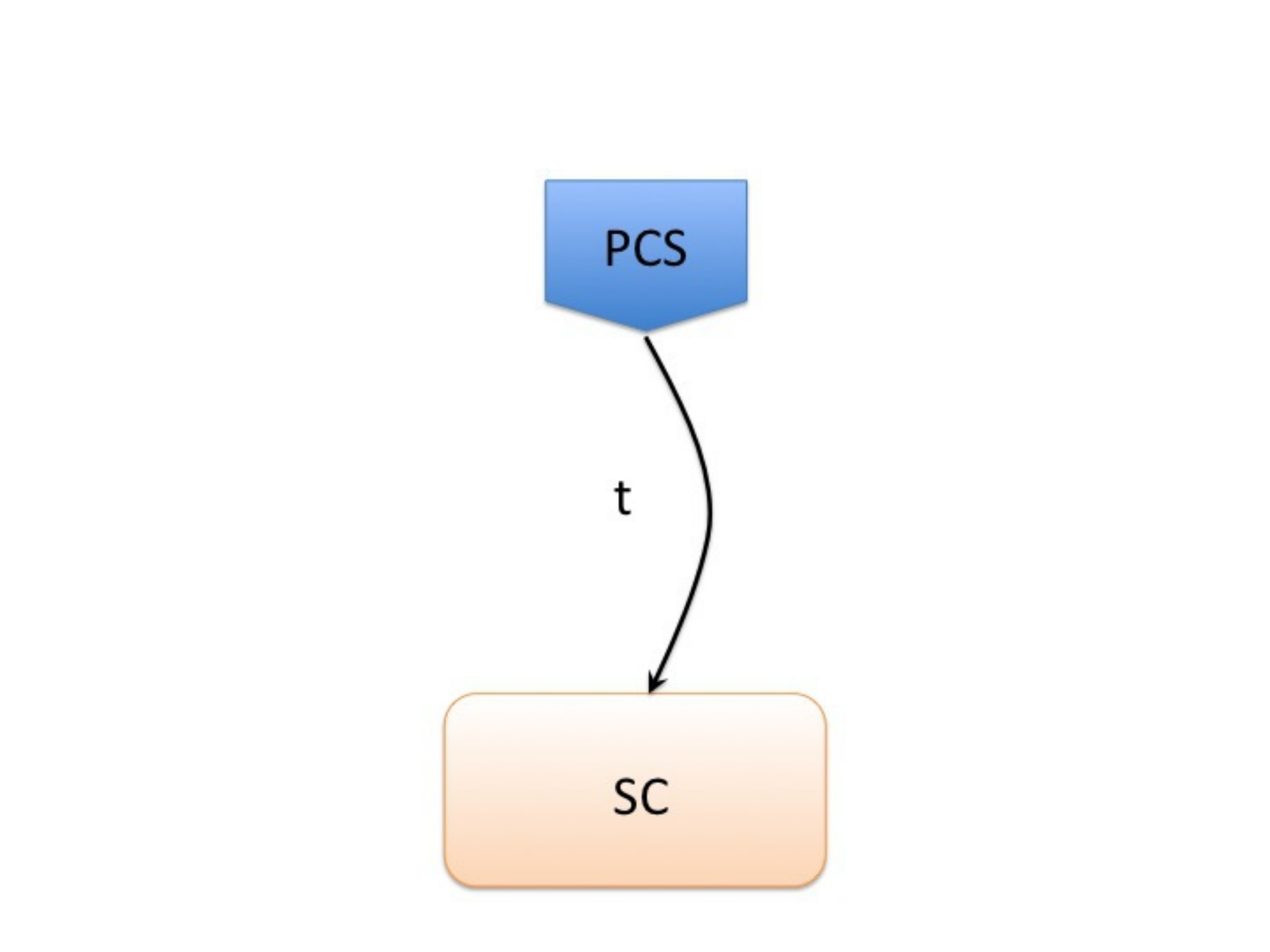}
\caption{(color online) Cartoon of point-contact tunneling process in the BTK regime with tunneling overlap integral $t$.
}
\label{fig:cartoon_btk}
\end{figure}

\begin{figure}[hbt]
\includegraphics[width=0.90\columnwidth,angle=0]{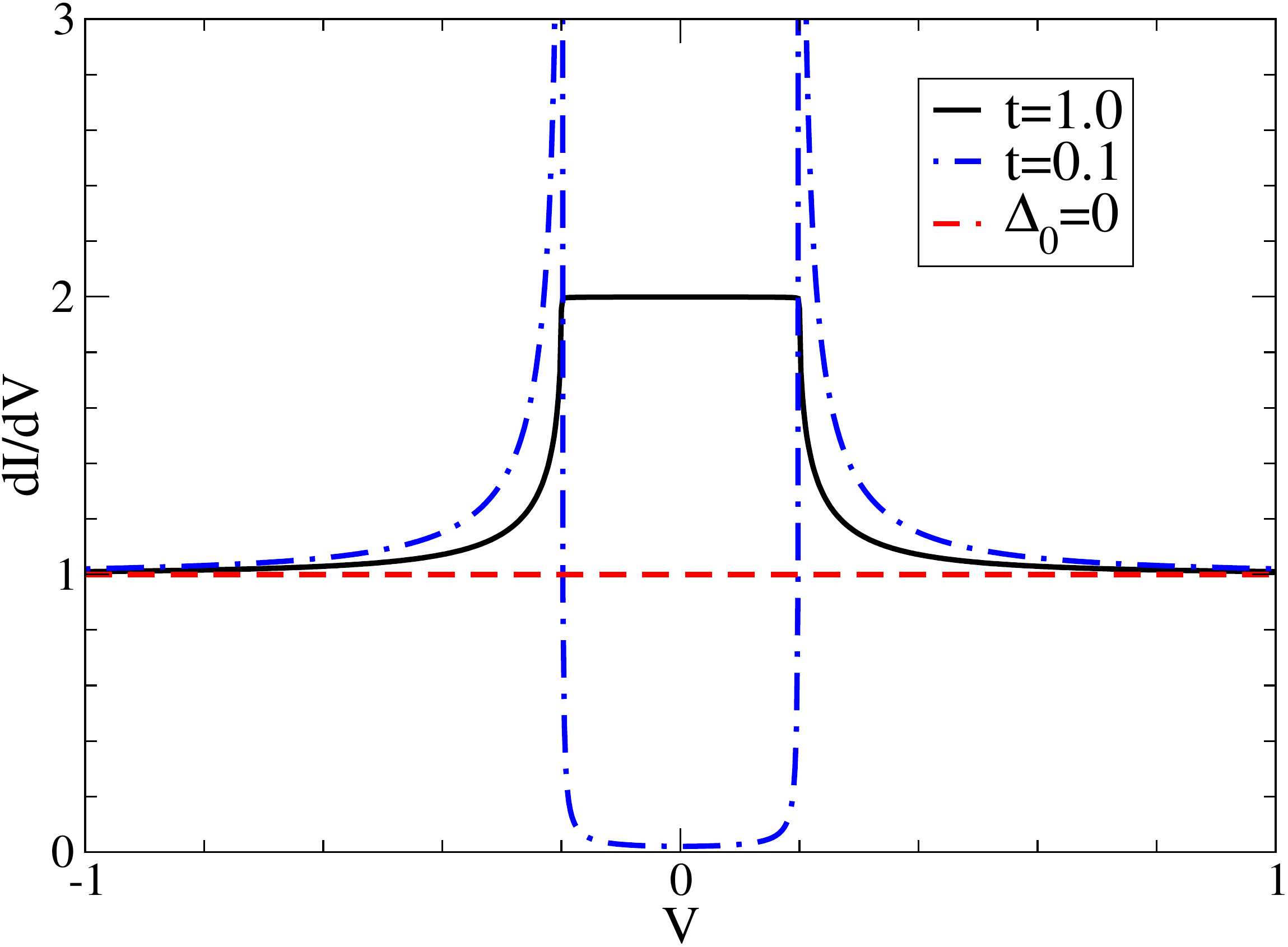}
\caption{(color online)  
The BTK differential conductance at $T=0$ is plotted for high ($t=1.0$) and low ($0.1$) transparency with $s$-wave gap $\Delta_0=0.2$  and in the normal state  with $\Delta_0=0$ (dashed red line).
}
\label{fig:fig_btk}
\end{figure}

For $t_{loc}=v=E_0=0$, the expression in Eq.~(\ref{eq:kernel_SC}) reduces to the standard conductance formula of BTK\cite{btk} and is equivalent to Eq.~(17) of Ref.~\onlinecite{Kashiwaya1996} (after setting their parameters $\theta_S=\theta_N=0$). The basic tunneling process is illustrated in Fig.~\ref{fig:cartoon_btk}.
The two extreme tunneling limits of low (${\cal D}\ll 1$ or $t\ll 1$) and high (${\cal D} = 1$ or $t= 1$) transparency are shown for an $s$-wave superconductor 
in Fig.~\ref{fig:fig_btk} 
as baseline for further comparison with the more general tunneling cases below.

\subsection{Lorentzian Line Shape Regime}

\begin{figure}[hbt]
\includegraphics[width=0.95\columnwidth,angle=0]{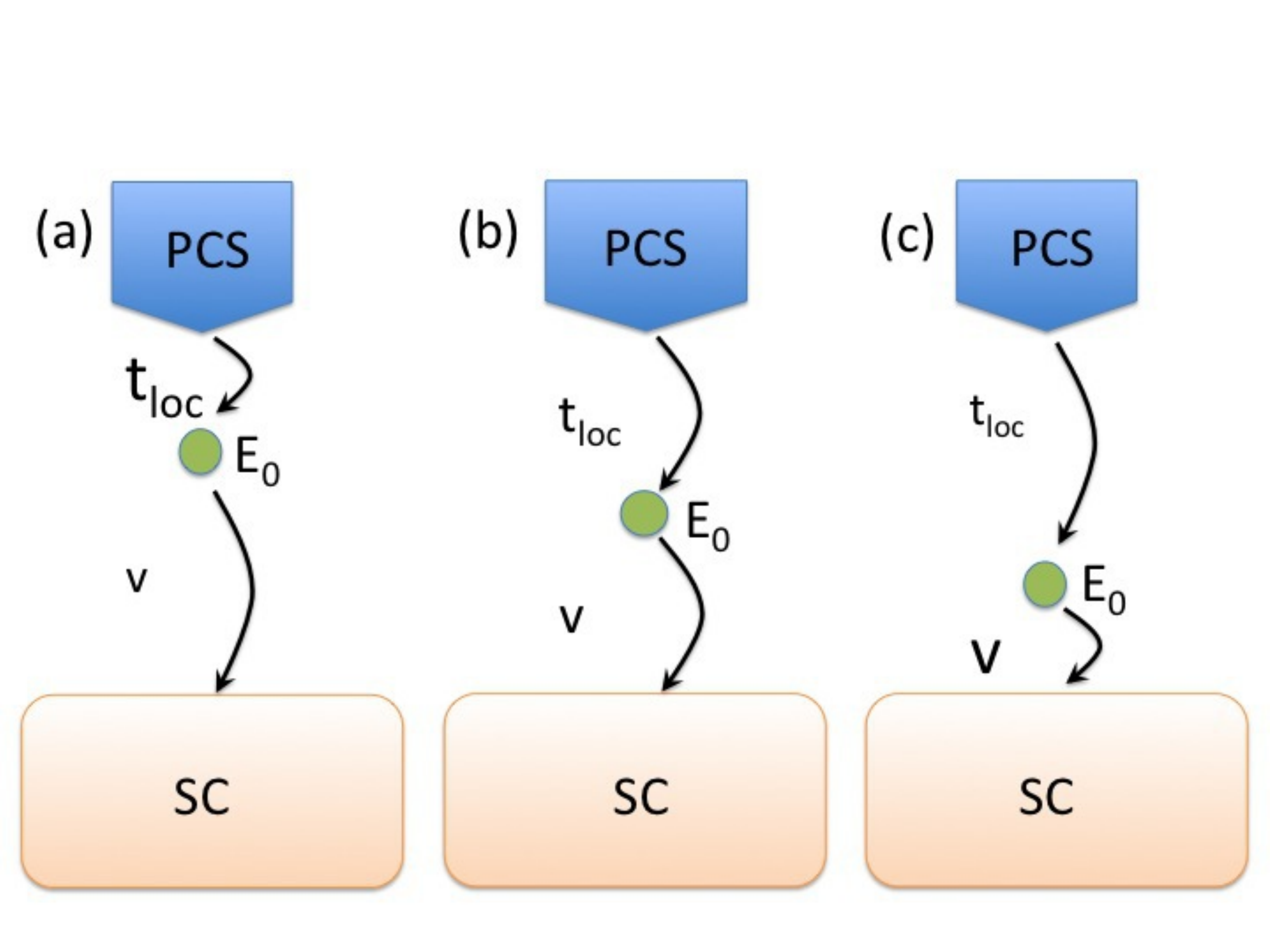}
\caption{(color online) Cartoon of PCS tunneling through the localized state with energy level $E_0$ 
into the superconductor (SC) for the Lorentzian line shape regime with $t=0$.
We show tunneling limits (a) $t_{loc}\gg v$, (b) $t_{loc}\sim v$, and (c) $t_{loc} \ll v$.
}
\label{fig:cartoon_lorentzian}
\end{figure}

\begin{figure}[hbt]
\includegraphics[width=0.95\columnwidth,angle=0]{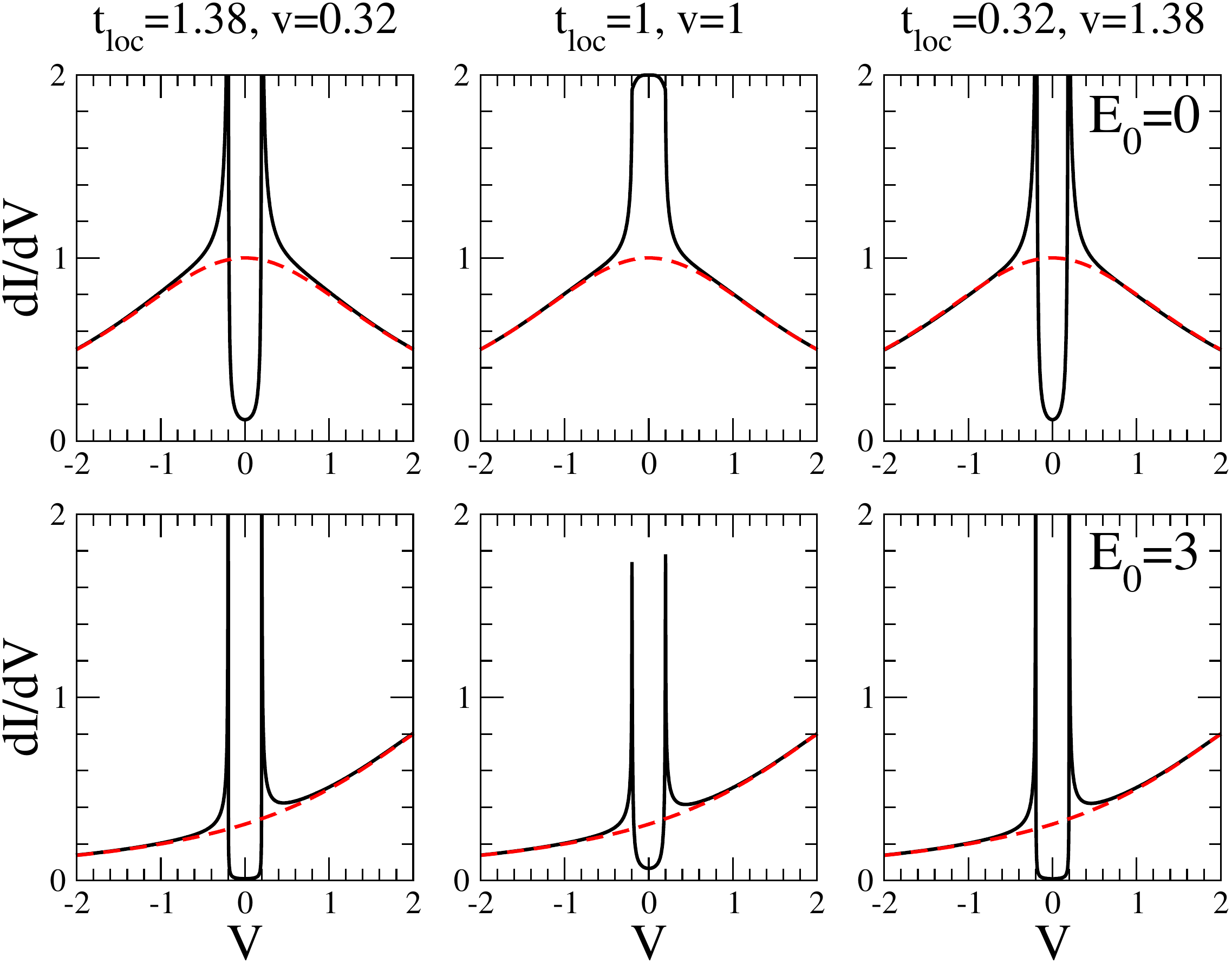}
\caption{(color online) The Lorentzian differential conductance at $T=0$ for resonant $E_0=0$ (top) and off-resonant $E_0=3$ (bottom) localized state with $s$-wave gap $\Delta_0=0.2$ (black solid line) and normal state $\Delta_0=0$ (dashed red line).
The columns depict from left to right three characteristic parameter sets,
$t_{loc} = 1.38, v=0.32$,
$t_{loc} = v =1.0$, and
$t_{loc}=0.32, v=1.38$. 
}
\label{fig:fig_lorentz}
\end{figure}

For $t=0$, the expression in Eq.~(\ref{eq:kernel_SC}) reduces to tunneling through a localized state  into the superconductor with a Lorentzian line shape of the differential conductance. The localized state can be either an adatom (impurity) of the superconductor or the metal contact or in between. This scenario is similar to tunneling into a single nonmagnetic impurity on a metal surface, which has been studied before with scanning tunneling spectroscopy.\cite{Crommie1993}
The basic tunneling processes are illustrated in Fig.~\ref{fig:cartoon_lorentzian},
where three limits  
(a) $v/t_{loc} \ll 1$ (adatom or impurity atom on contact),
(b) $v=t_{loc}$ (impurity between), and
(c) $v/t_{loc} \gg 1$ (impurity on SC) are shown. 
In contrast to the BTK tunneling conductance curves
in Fig.~\ref{fig:fig_btk}, we see in Fig.~\ref{fig:fig_lorentz} the effects of the Lorentzian line shape for
both resonant ($E_0=0$) and off-resonant ($E_0 =3 > \Gamma$) tunneling into an $s$-wave superconductor.  
For resonant tunneling  the differential conductance is maximum and symmetric around the zero voltage bias.
The main results of the Lorentzian line shape are that (1) the position of the localized state cannot be differentiated 
between the impurity close to the PCS contact or close to the superconductor and (2) that
significant Andreev reflection (100\%) is only possible for close to resonant tunneling, $E_0=0$,
when $t_{loc}\sim v$.
Otherwise the differential conductance curves exhibit the low-transparency or tunneling limit with the BCS coherence peaks
at the gap edge $\Delta_0$.

We find by plotting the conductance curves in Fig.~\ref{fig:fig_lorentz} that the broken symmetry in the superconducting state, due to the exchange of $t_{loc}\leftrightarrow v$, is most likely too subtle to be detected within the range of parameters. The line shapes of the conductance in the left and right columns of Fig.~\ref{fig:fig_lorentz} are nearly indistinguishable, and hence would require high-precision measured PCS conductance curves to assign with confidence the localized state to an impurity on either the tip or sample.

\subsection{Fano Line Shape Regime}

\begin{figure}[hbt]
\includegraphics[width=0.95\columnwidth,angle=0]{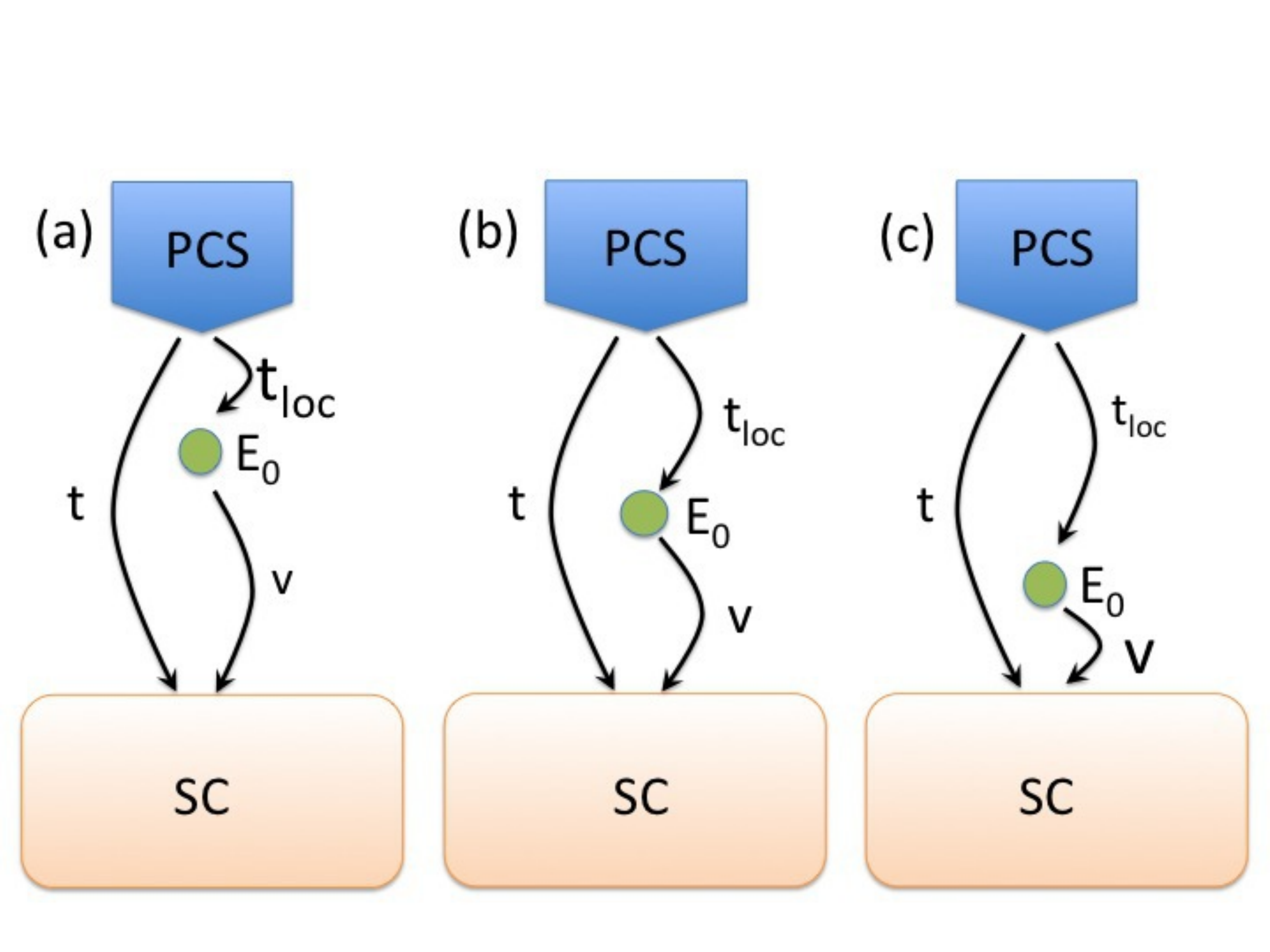}
\caption{(color online) Cartoon of PCS tunneling through the localized state with energy level $E_0$
into the superconductor (SC) for the Fano line shape regime with $t\ne 0$.
We show tunneling limits (a) $t_{loc}\gg v$, (b) $t_{loc}\sim v$, and (c) $t_{loc} \ll v$.
}
\label{fig:cartoon_fano}
\end{figure}

\begin{figure}[hbt]
\includegraphics[width=0.95\columnwidth,angle=0]{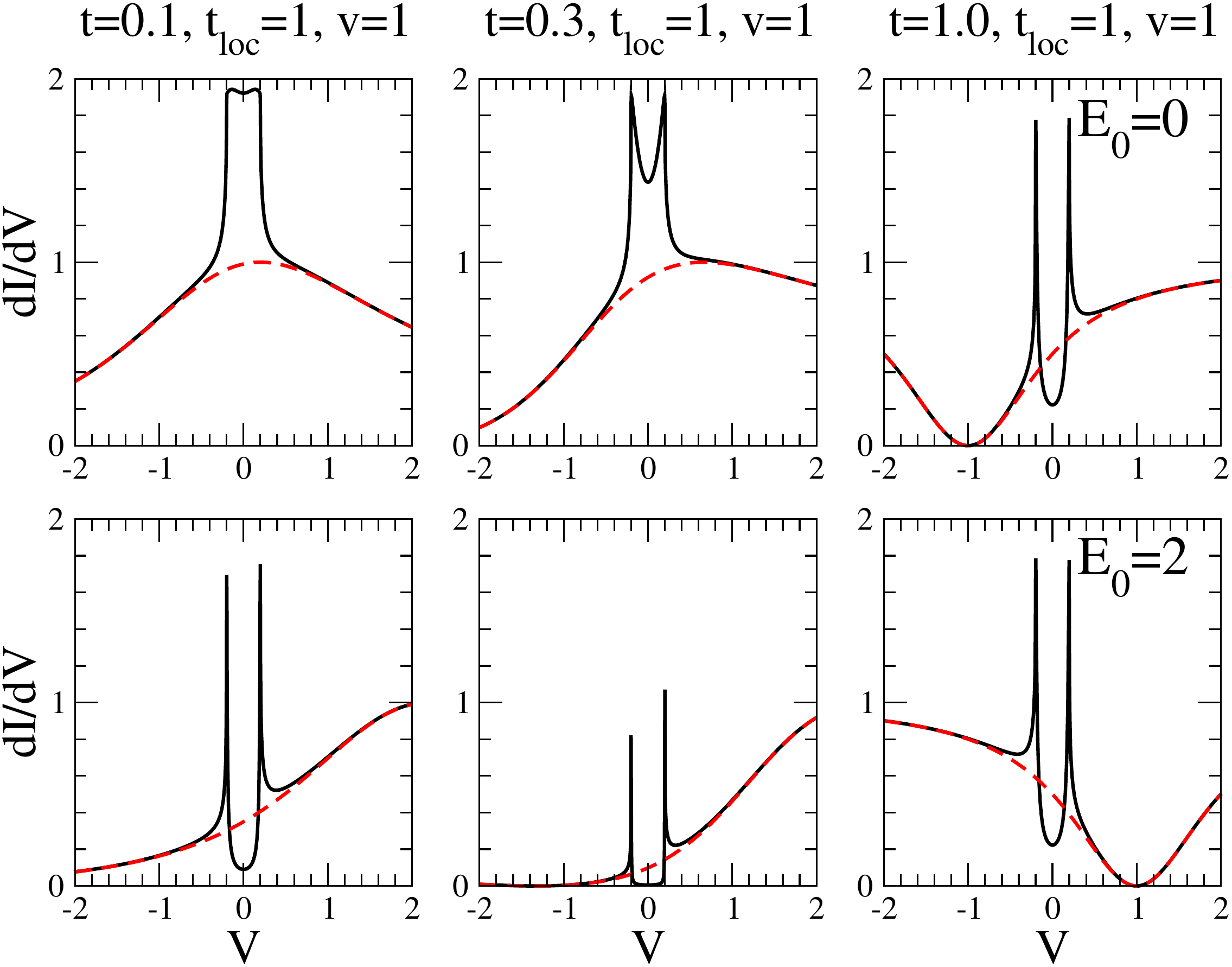}
\caption{(color online) The Fano differential conductance at $T=0$ for resonant $E_0=0$ (top) and off-resonant $E_0=3$ (bottom) localized state with $s$-wave gap $\Delta_0=0.2$ (black solid line) and normal state $\Delta_0=0$ (dashed red line).
The columns depict three characteristic parameter sets with increasing direct tunneling $t$ from left to right,
$t=0.1, 0.3, 1.0$ and  
$t_{loc} = v=1.0$ otherwise.
}
\label{fig:fig_fano}
\end{figure}

The Fano resonance arises from the quantum mechanical interplay between interfering tunneling paths via the localized state and a continuum of itinerant states.
Here we allow all model parameters to vary. 
However, we limit our discussion of Fano tunneling to the most relevant cases for PCS measurements in HFSs.
This scenario is similar to the Kondo resonance, which has been studied by tunneling into a single magnetic impurity on a metal surface.\cite{Madhavan1998}
In Fig.~\ref{fig:fig_fano}, we see 
that
significant Andreev reflection (100\%) is only possible for close to resonant tunneling
and when direct tunneling is weak compared to tunneling via the localized state $t < t_{loc}\sim v$.
Otherwise the differential conductance curves are dominated by the line shape of the Fano resonance with
asymmetric BCS coherence peaks at $\Delta_0$.

In a series of theoretical works \cite{Maltseva2009, Yang2009, Woelfle2010, Figgins2010, Yuan2012}
the question of the observed Fano line shape in $dI/dV$ tunneling spectra of heavy fermions URu$_2$Si$_2$ and \cecoin\
was addressed.\cite{Schmidt2010, Aynajian2010, Aynajian2012,parkwk2012, Allan2013}
In particular, W\"olfle and coworkers argued that hybridization between itinerant conduction electrons and localized $f$ electrons will always generate a hybridization gap and that the $dI/dV$ characteristic will not show a Fano line shape, unless strong correlations broaden the heavy quasiparticle states. In that case, the inclusion of an electron self-energy fills in the gap and results in a Fano-like line shape similar to a Kondo impurity as observed in the scanning tunneling experiments.

On the other side, Yang \cite{Yang2009} used the slave-boson mean-field approximation to calculate the PCS conductances of \cecoin\ and \cerhin.
Since he assumed a constant density of states for both electrons in the normal metal tip and in the hybridized light and heavy bands of the heavy-fermion compound, he circumvented the hybridization gap dilemma and found a Fano resonance. However, in order to fit the experimental PCS conductances he then was forced to invoke a complex Fano factor with unphysically large imaginary part. This 
model fit motivated us to construct a minimal PCS tunneling model that shows both a Fano line shape and on top of that Andreev reflection in the superconducting state.

\subsection{Soft PCS Measurements}

In recent years it has been demonstrated that the soft point-contact spectroscopy 
can be adapted to a high-pressure environment.\cite{Lu2012, Sakai2012}
Instead of the conventional PCS method with a sharp metal tip, SPCS contacts used here are made by dipping the
end of a 25 $\mu{\rm m}$-diameter platinum wire into Ag epoxy and attaching it to the 
[001] surface of the crystal.
This method has been successfully implemented for the study of the superconducting order parameter of various
superconductors  at ambient pressure (see the review Ref.~\onlinecite{Daghero2010} and references within),
as well as the hidden order and antiferromagnetic phases of URu$_2$Si$_2$ at high pressure, \cite{Lu2012, Sakai2012}
with the advantage of reliable stability over a large temperature range. 

Here we extend the SPCS  technique
to  study the heavy-fermion superconductor \cecoin\ at ambient pressure as well as perform pressure studies of \cerhin. 
The single crystals 
were mounted in a teflon capsule of a clamped toroidal pressure cell, filled with glycerine-water fluid (3:2)  as pressure transmitting medium, which provides
a very nearly hydrostatic environment. 
The pressure at low temperature was determined from the resistively measured
change in the superconducting transition temperature of Pb.
The differential conductance $G = dI/dV$ as a function of bias voltage $V$ was recorded by a standard lock-in technique,
with the sample biased positively for all the measurements.

The superconducting transition temperatures of the single crystals measured are $T_c=2.3$ K for \cecoin, 
and $T_c\sim 2$ K for both \cerhin\ at 1.8 and 2.0 GPa.

A visual inspection of the asymmetric line shape of the SPCS conductances in Figs.~\ref{fig:CeCoIn5_NS} and
\ref{fig:CeRhIn5_1.8GPa_NS}
suggests that the point contacts of both crystals are in the ballistic regime. In the case of \cecoin\ the in-plane resistivity is $\rho\sim 3\, \mu\Omega$cm and the mean-free path was estimated as $\ell\sim 81$ nm,\cite{Movshovich2001}
whereas for \cerhin\ the in-plane resistivity is $\rho\sim 1\, \mu\Omega$cm at 1.8 and 2.0 GPa.\cite{Park2011}  
Lacking an estimate for the mean-free path in \cerhin, we assume the same value as for \cecoin. 
Finally, we can verify that the contacts are in the ballistic regime by using the definition of the Sharvin resistance $R_S = 16\rho\ell/(3\pi d^2)$, \cite{Sharvin1965}
which is given by $R_S=1/G_0$.
The extracted diameters of the PCS contacts are
$d=34$ nm for \cecoin, $d=19$ nm for \cerhin\ at 1.8 GPa, and $d=39$ nm at 2.0 GPa. 
In summary, for all cases the ratio $(d/\ell)^2 \ll 1$ is consistent with the assumption of ballistic contacts.
In particular, the characterization of our SPCS contact for \cecoin\ is in good agreement with conventional PCS measurements by Park et al.,\cite{parkwk2005}
who reported the Sharvin limit with an estimated upper size of $d=46$ nm.

\subsection{Normal State Analysis}\label{subsection_ns}

\begin{table}[b]
\caption{Fitting the phenomenological model parameters to the SPCS curves.
Data for \cecoin\ are obtained at ambient pressure and $T=1.31$ K;
\cerhin\ at $P=1.8$ GPa and $T=1.16$ K, and for $P=2.0$ GPa and $T=1.18$ K.
For \cecoin\ and \cerhin\ at 2.0 GPa the background $G_0(V)$ was modeled to be constant, while for \cerhin\ at 1.8 GPa the functional dependence
$G_0(V)=G_0-G_1 \tanh{(V/V^*)}$ was assumed.}
\begin{center}
\begin{tabular}{|c|c|c|c|c|c|c|c|c|}
\hline
SPCS 		&  $P$ & $ E $ 	& $\Gamma$	& $q_F$	& $C_0$ 					& $G_0$ 					& $G_1$&	$V^*$	\\
	   		& GPa  & meV 		& meV			&		& $1/({\rm k}\Omega)$ 	&  $1/({\rm k}\Omega)$  & $1/({\rm mV}\, {\rm k}\Omega)$ & mV\\
\hline
CeCoIn$_5$ &  $\sim 0$ & 1.1	& 6.9	&	-1.9	&	21.7				& 282.6				& 0		& -	\\ 
\hline
CeRhIn$_5$ &  1.8 & 5.4			& 8.0	& -1.7	&	3.70				& 261.4				& 11.0	& 55.0 \\      
\hline
CeRhIn$_5$ &  2.0 & 5.8			& 14.0	& -1.8	&	24.9				& 1117.1				& 0		& - \\   
\hline
\end{tabular}
\end{center}
\label{table:default}
\end{table}

\begin{figure}[th]
\includegraphics[width=0.95\columnwidth,angle=0]{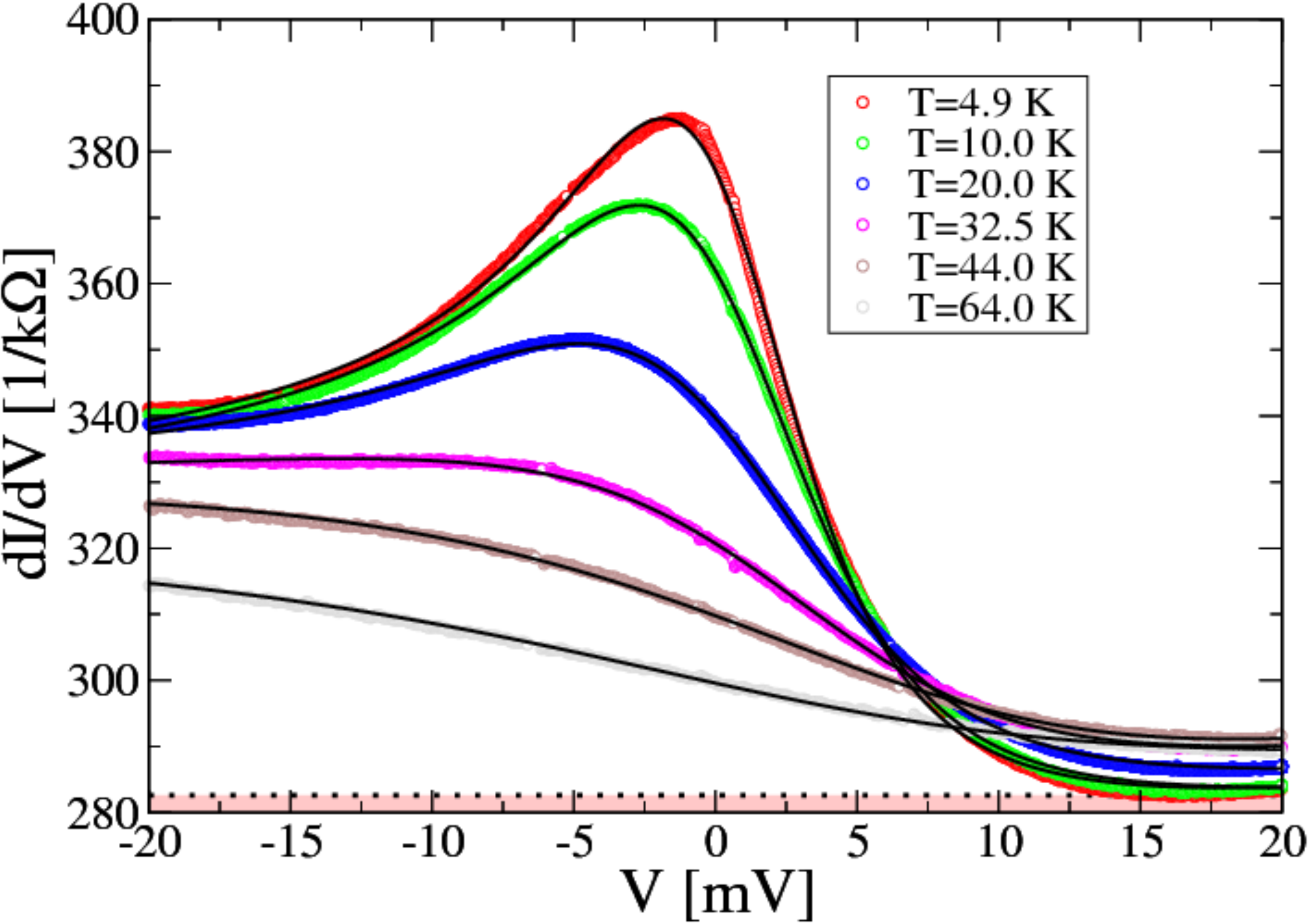}
\caption{(color online) Fitting the Fano line shape of \cecoin\ at ambient pressure in the normal state.
The transparency ${\cal D}=0.210$ is in the  tunneling rather than high transparency limit.
The solid (black) lines are fits with constant background conductance $G_0=282.6\, (k\Omega)^{-1}$,
which is shown as red-shaded background.
The contact's pin code is $(t, t_{loc}, v, E_0)=$
(0.243, 0.066 $\sqrt{\rm eV}$, -0.055 $\sqrt{\rm eV}$,-0.53 meV).
}
\label{fig:CeCoIn5_NS}
\end{figure}

\begin{figure}[bh]
\includegraphics[width=0.95\columnwidth,angle=0]{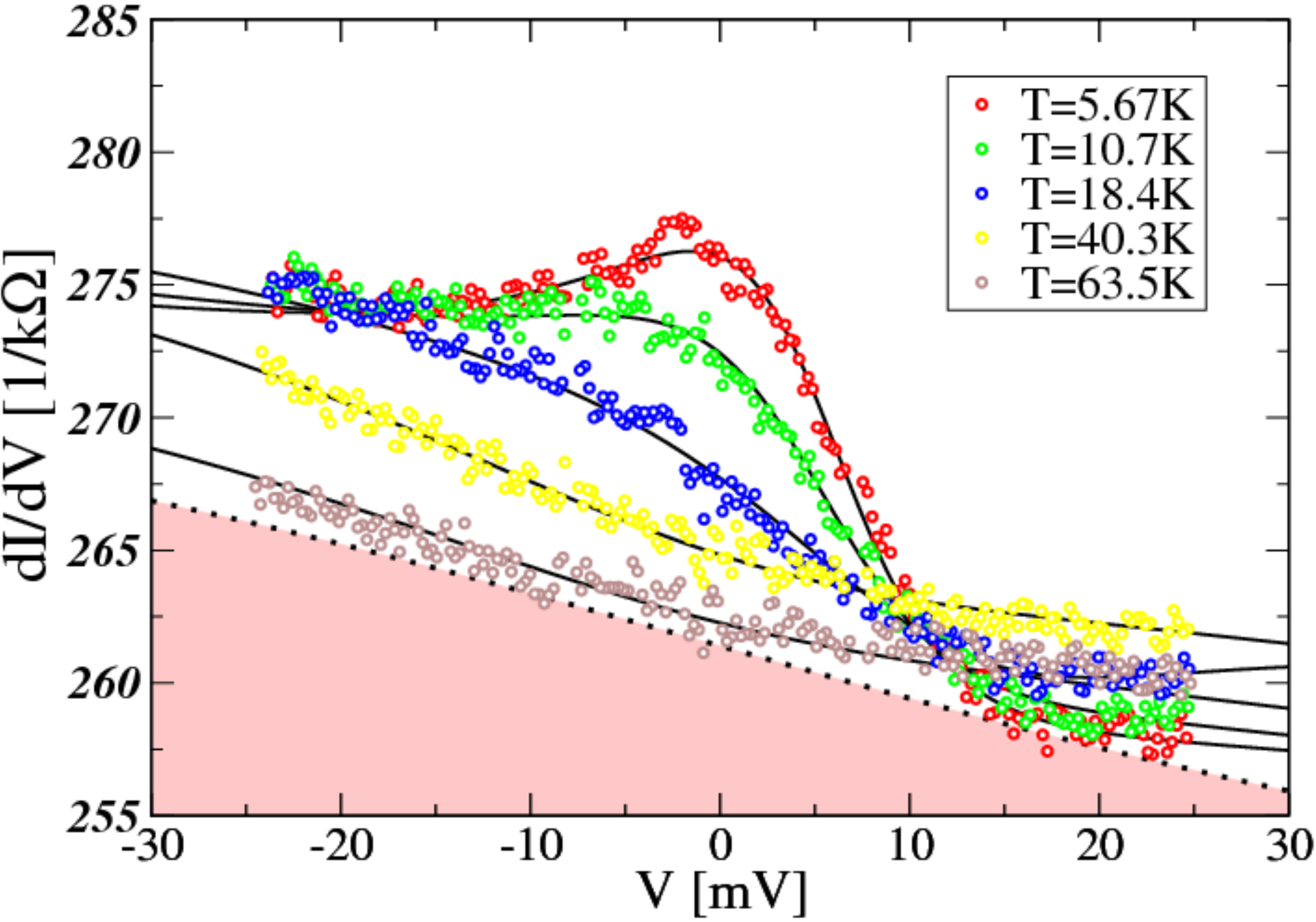}
\caption{(color online) Fitting the Fano line shape of \cerhin\ at 1.8 GPa in the normal state.
The transparency  ${\cal D}=0.532$ is in the intermediate transparency limit.
The solid (black) lines are fits with background conductance $G_0(V)$,
which is shown as red-shaded background.
The contact's pin code is 
(0.270, 0.066 $\sqrt{\rm eV}$, -0.065 $\sqrt{\rm eV}$, 3.29 meV).
}
\label{fig:CeRhIn5_1.8GPa_NS}
\end{figure}

We begin by fitting the phenomenological model parameters of the SPCS data sets for a series of different temperatures to extract the temperature-independent background function $G_0(V)$. 
A simple and plausible choice is $G_0(V)=G_0-G_1 \tanh{(V/V^*)}$.
While an asymmetric background conductance is rare to occur in point-contact measurements with conventional metals,\cite{Naidyuk1985} 
it has also been reported for 
other correlated electron systems like the high-temperature iron-based \cite{Mehta2010, Tortello2012, Naidyuk2014, Arham2014} 
and cuprate superconductors.\cite{Murakami2000, Anderson2006} 

The phenomenological model parameters for \cecoin\ and \cerhin\ are given in Table.~\ref{table:default} and the corresponding spectra are shown in 
Figs.~\ref{fig:CeCoIn5_NS} and ~\ref{fig:CeRhIn5_1.8GPa_NS}.
For each individual $dI/dV$-curve we are able to obtain a good fit to a Fano resonance over the entire measured voltage window. 
However, we note that the $dI/dV$-curves of \cerhin\ at 1.8 GPa are only well fit within the chosen voltage window 
when a voltage-dependent background conductance $G_0(V)=G_0-G_1 \tanh{V/V^*}$ is assumed with 
fit parameters $G_1=11.0$ (mV \, k$\Omega$)$^{-1}$ and a rather large $V^*=55.0$ mV.
Alternatively, a simple linear background function would fit the data as well, as can be seen by the red-shaded background curve
in Fig.~\ref{fig:CeRhIn5_1.8GPa_NS}.

From this fit procedure we find that the Fano parameter $q_F$ and the line broadening
$\Gamma$ are similar for both materials, while the renormalized energy levels $E$ of the surface state differ significantly.
Quantitatively similar parameters were obtained for \cecoin\ with conventional PCS  metal tips analyzed in Ref.~\onlinecite{Fogelstrom2010}.
In contrast, the PCS data by Rourke et al.\cite{Rourke2005} for \cecoin\  were most likely not in the ballistic regime, which would explain the significantly different line shape of their $dI/dV$ curves.
Assuming a constant background conductance for \cecoin,  we found in Ref.~\onlinecite{Fogelstrom2010}  at $T\sim 5$ K for a gold tip:
($E\sim 2$ meV, $\Gamma\sim 13$ meV, $q_F\sim -2.1$, $C_0\sim 5.6$ 1/k$\Omega$, $G_0\sim 163$ 1/k$\Omega$),
and for a platinum tip:
($E\sim 2.5$ meV, $\Gamma\sim 17$ meV, $q_F\sim -1.8$, $C_0\sim 2.0$ 1/k$\Omega$, $G_0\sim 41$ 1/k$\Omega$),
for details see Table~I  and Fig.~5 of Ref.~\onlinecite{Fogelstrom2010}.
The  origin of the larger zero-bias background conductance $G_0\sim 282$ $1/k\Omega$ for SPCS contacts compared to 163 or 41 $1/k\Omega$ for conventional PCS contacts is unknown, but could be due to  a larger contact area of the micro-meter-sized particles in the silver paint, which is also expressed in the larger values of $C_0$.

\begin{table}[htb]
\caption{The pin codes of the soft point contacts.
The uniqueness of the microscopic parameters was determined by fitting to the Andreev signal in the superconducting state, since in the normal state the parameters $t$ and $E_0$ are correlated.}
\begin{center}
\begin{tabular}{|c|c|c|c|c|c|}
\hline
SPCS 		&  $P$  		& $ t $ 	& $t_{loc}$		& $v$	& $E_0$ \\
	   		& GPa  		&   		& $\sqrt{eV}$	& $\sqrt{eV}$	& meV 	 \\
\hline
CeCoIn$_5$ &  $\sim 0$ 	& 0.243	& 0.066			& -0.055 		& -0.53	 \\ 
\hline
CeRhIn$_5$ &  1.8 		& 0.270	& 0.066			& -0.065 		&  3.29   \\      
\hline
CeRhIn$_5$ &  2.0 		& 0.253 & 0.096			& -0.075		& 2.36   \\   
\hline
\end{tabular}
\end{center}
\label{table:pincode}
\end{table}

In the following analysis of point-contact differential conductances, we introduce a {\it pin code} to describe each SPCS fit at the lowest measured temperature. 
Ideally the pin code is a unique sequence of tunneling model parameters
$(t, t_{loc}, v, E_0)$ 
characteristic of each point-contact tunnel junction.
A similar pin code scheme was introduced to characterize the number of current carrying channels of one-atom sized contacts.\cite{Scheer1998}
From the phenomenological model parameters in Table~\ref{table:default} we extract at each temperature the microscopic model parameters in form of overlap integrals
and the energy level of the localized state
by using Eqns.~(\ref{eq:parameters_begin})-(\ref{eq:parameters_end}). 
Unfortunately, in the normal state the microscopic parameters $t$ and $E_0$ are correlated, see e.g.\ Fig.~7 in Ref.~\onlinecite{Fogelstrom2010}. A unique determination is only possible by fitting the Andreev signal in the superconducting state.
The unique pin codes
of our samples are given in 
Table~\ref{table:pincode} 
with the corresponding phenomenological fit parameters in Table~\ref{table:default}.
While the pin codes are quite similar for all three cases, we cannot discern a simple trend with applied pressure or between \cecoin\ and \cerhin.
For example, we have no explanation in terms of the bulk heavy-fermion state to why the tunneling parameter $t$ decreases in \cerhin\ from 1.8 to 2.0 GPa, while $t_{loc}$ and $v$ increase, or why the localized state $E_0$ in \cecoin\ lies below the Fermi level in contrast to \cerhin.
On the other hand, these apparent random changes in microscopic tunneling model parameters corroborate our hypothesis that localized surface states cause the Fano line shape in the $dI/dV$-curves. Whenever a point contact is formed surface defects are created randomly leading to the scatter in model parameters observed.

\subsection{Superconducting State Analysis}\label{subsection_sc}

\begin{figure}[b]
\includegraphics[width=0.95\columnwidth,angle=0]{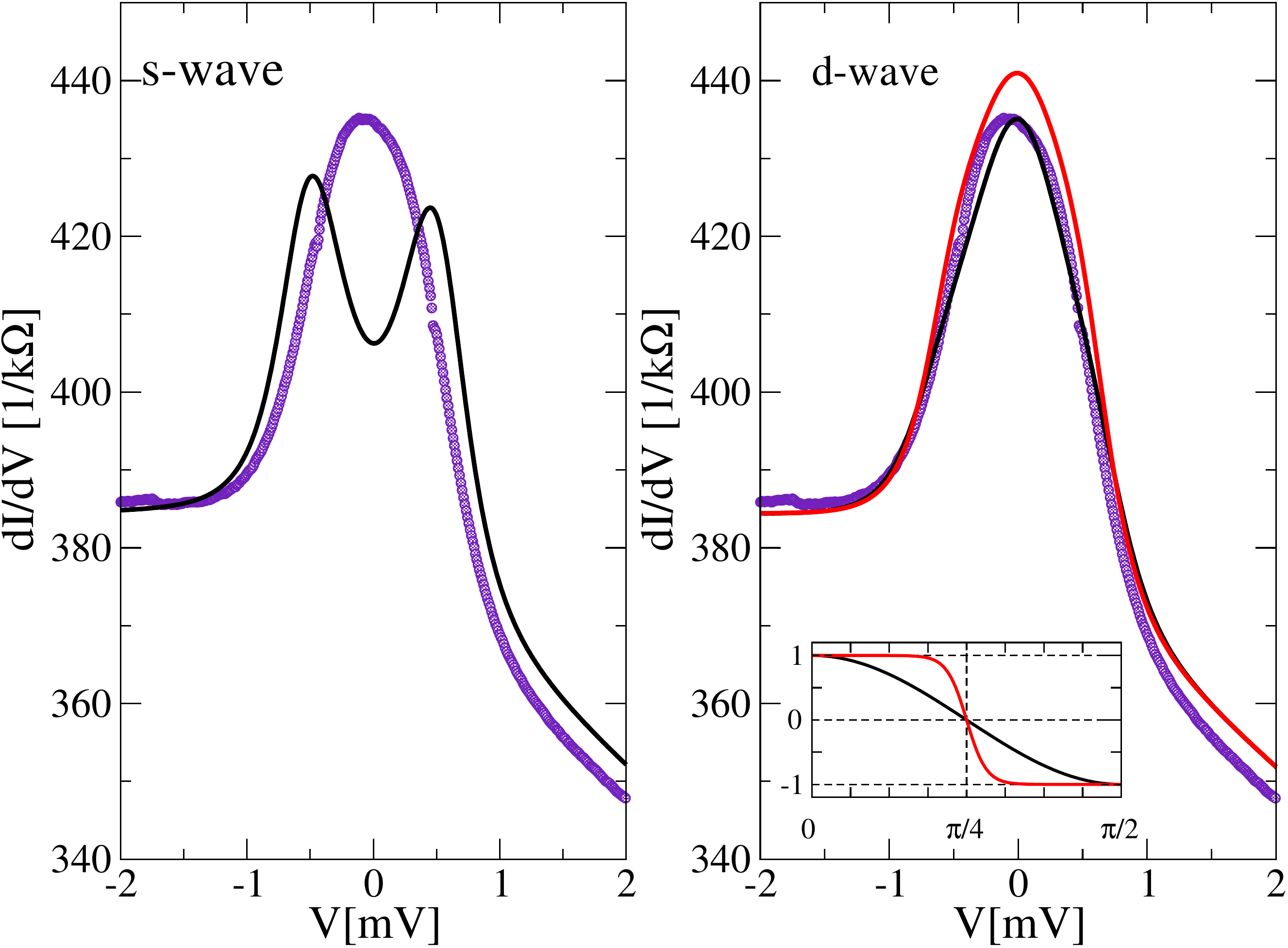}
\caption{(color online) Fitting the Andreev signal of \cecoin\ in the superconducting state at ambient pressure and at 1.31 K with $T_c=2.3$ K.
We assumed isotropic $s$-wave gap (left) and $d$-wave gap (right).
The $d$-wave is plotted for standard (black) and five times steeper (red)
slope of the gap at nodes  
with equal weight of  tunneling into [100] ($0^\circ$) and [110] ($45^\circ$)  surface orientations.
Inset: Angle dependence of gap function.
}
\label{fig:CeCoIn5_SC}
\end{figure}

\begin{figure}[th]
\includegraphics[width=0.95\columnwidth,angle=0]{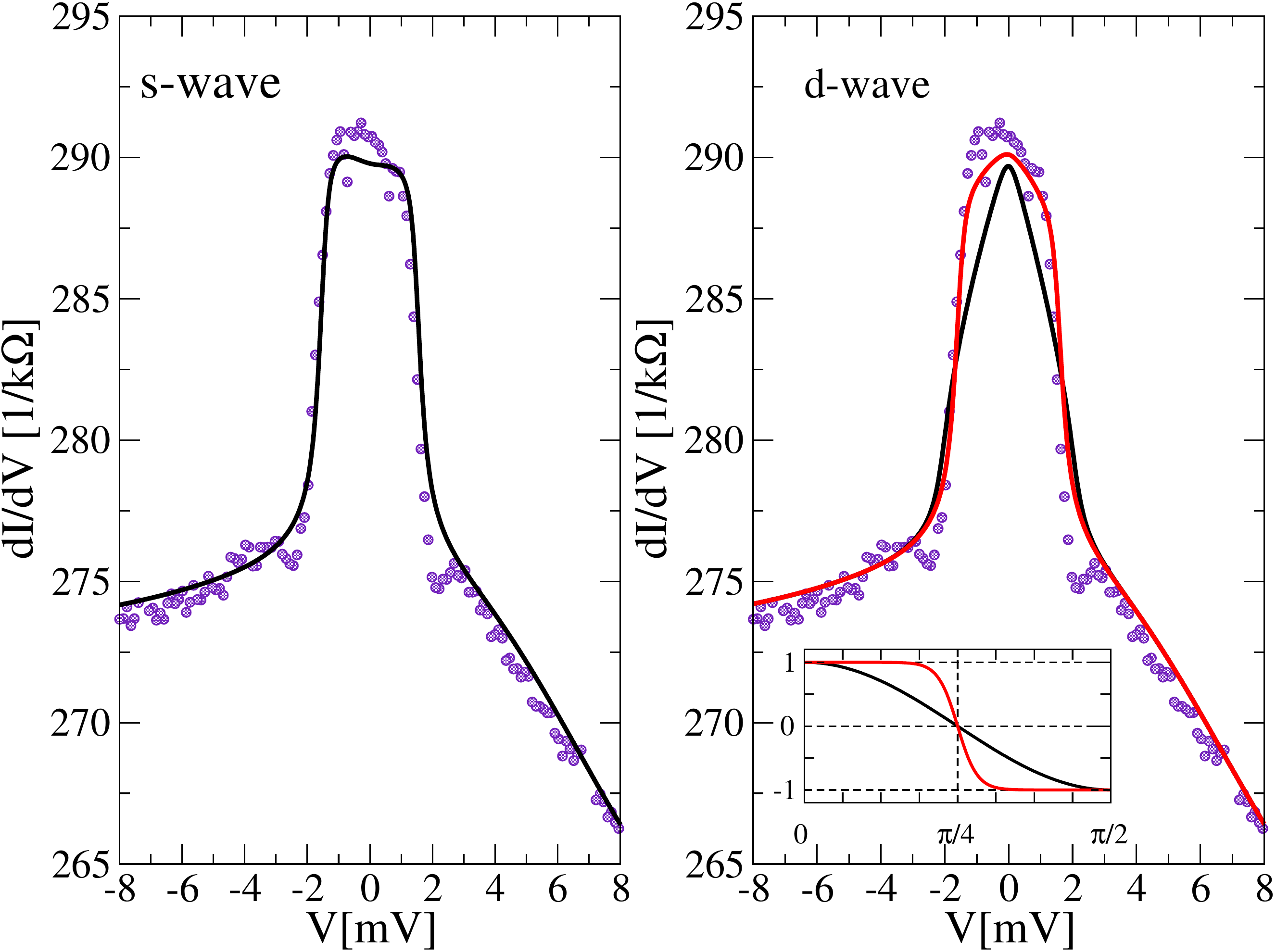}
\caption{(color online) Fitting the Andreev signal of \cerhin\ in the superconducting state at 1.8 GPa and at 1.16 K  with $T_c\sim 2.0$ K
and $\Delta_0=$1.76 meV.
We assumed isotropic $s$-wave gap (left) and $d$-wave gap (right).
The $d$-wave is plotted for standard (black) and five times steeper (red)
slope of the gap at nodes  
with equal weight of  tunneling into  [100] and [110] surface orientations.
Inset: Angle dependence of gap function.
}
\label{fig:CeRhIn5_1.8GPa_SC}
\end{figure}

\begin{figure}[th]
\includegraphics[width=0.95\columnwidth,angle=0]{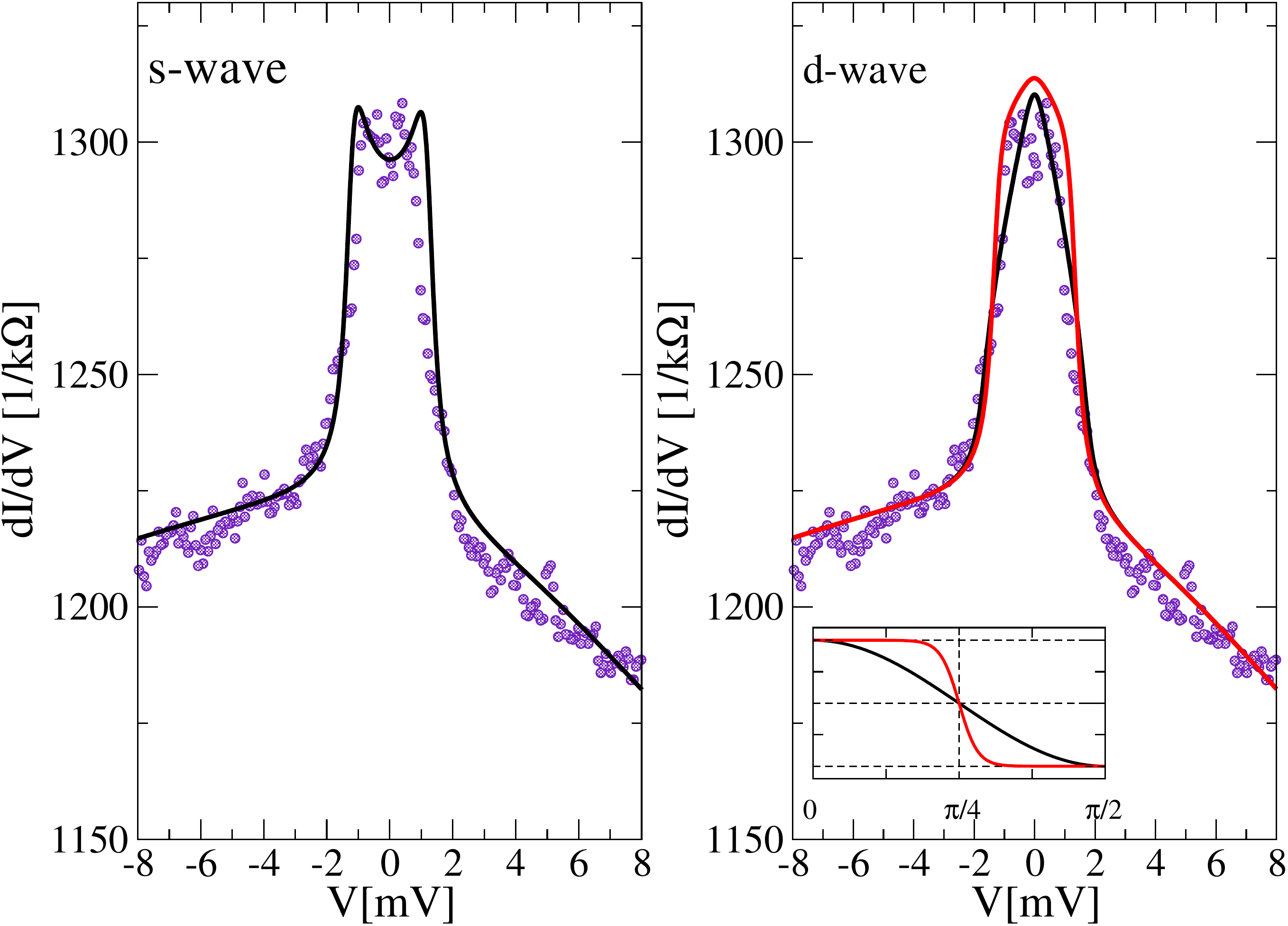}
\caption{(color online) Fitting the Andreev signal of \cerhin\ in the superconducting state at 2.0 GPa  and at 1.18 K with $T_c\sim 2.0$ K
and $\Delta_0=$1.38 meV.
The contact's transparency is ${\cal D}=0.253$, which is in the  tunneling rather than high transparency limit.
We assumed isotropic $s$-wave gap (left) and $d$-wave gap (right).
The $d$-wave is plotted for standard (black) and five times steeper (red)
slope of  the gap at nodes  
with equal weight of  tunneling into  [100] and [110] surface orientations.
The contact's pin code is 
(0.253, 0.096 $\sqrt{\rm eV}$, -0.075 $\sqrt{\rm eV}$, 2.36 meV).
Inset: Angle dependence of gap function.
}
\label{fig:CeRhIn5_2.0GPa_SC}
\end{figure}

In the superconducting state one can use the nonlinear voltage dependence of the conductance to extract
further information about the remaining undetermined microscopic model parameter, e.g., 
${\cal D}$ (transparency) or $t$ (tunneling overlap integral).
This now uniquely determines all microscopic parameters of the two-channel tunneling model.

As already discussed in Ref.~\onlinecite{Fogelstrom2010}, the PCS conductance
is sensitive to the transparency ${\cal D}$ and the position of the localized state relative to the Fermi level.
Tunneling through the resonant state, $E_0=0$,
enhances the effective transparency of the junction, so that a contact with ${\cal D}\ll 1$ has a conductance similar to the BTK conductance of transparency ${\cal D}\sim 1$. Another crucial result of the
two- or multichannel models is that the conductance enhancement due
to Andreev reflection can be tuned to only a few percent relative to the normal-state conductance versus
the conventional 100\% of the single-channel BTK model. Note that the suppression
of the Andreev reflection signal in the HFS comes naturally about by including the poorly understood background conductance $G_0(V)$.

In Fig.~\ref{fig:CeCoIn5_SC} we see good agreement between the measured Andreev reflection signal and our calculation for \cecoin\ assuming $d_{x^2-y^2}$-wave pairing with equal mixture of tunneling into [100] ($0^o$) and [110] ($45^o$) interface orientations. 
An $s$-wave gap cannot describe the measured differential conductance, nor can a $d$-wave with a five times steeper gap slope. The angle dependence of the $d$-wave gap is shown in the inset.
The fitted gap amplitude of $\Delta_0=0.6$ meV of the standard $d$ wave is in excellent agreement with conventional PCS results.\cite{Fogelstrom2010}

The equal-mixture approximation for tunneling into the [100] and [110] crystal orientations needs some further explanations.
First, the conductance line shape is inconsistent with dominantly tunneling along the [001] direction in the low transparency limit, see Fig.~\ref{fig:fig_btk}.
Second, even with the soft PCS method a metal tip or silver particle is pressed into the sample, thereby creating sideway tunneling channels along all possible in-plane interfaces. 
Third, since neither pure [100] nor [110]  conductance calculations agree with the $dI/dV$ curves (not shown), we performed a minimalist's average over all possible interface orientations by averaging only the two extreme cases. For the  [100] orientation no pairbreaking of the superconducting order parameter occurs at the interface, while pairbreaking is maximal for the [110] orientation.
Of course, more realistic tunneling models would have to include an average over all possible orientations as well as the restricted size of the tunneling cone. Since our simplifications already result  in good fits, we do not expect to see much quantitative improvement by incorporating these details.

The situation is different for \cerhin. 
In Fig.~\ref{fig:CeRhIn5_1.8GPa_SC} we show at the pressure of 1.8 GPa and at the temperature of $T=1.16$ K ($T_c\sim 2$ K)
a somewhat 
flatter Andreev reflection signal with large gap $\Delta_0=1.76$ meV.
The plateau-like shape of the $dI/dV$ curve appears to be
more consistent with an isotropic $s$-wave gap 
or modified $d$-wave gap 
with a five times steeper gap opening than the standard $d_{x^2-y^2}$-wave gap function.
The angle dependence of the $d$-wave  gap is shown in the inset.
Such a modified $d$ wave has a reduced nodal quasiparticle region, but has the advantage that it still results in low-temperature power laws compared to the fully gapped $s$-wave scenario.
We speculate that the reason for this Andreev reflection signal is due to the proximity to 
a quantum critical state at this pressure.
At the pressure of 2.0 GPa and at $T=1.18$ K ($T_c\sim 2$ K), where we are now deeper into the SC dome, we see 
in Fig.~\ref{fig:CeRhIn5_2.0GPa_SC} that
the Andreev reflection signal is 
rather more peaked, though there is more scatter in the data around the zero-voltage bias.
Given the scatter in the conductance the Andreev reflection signal is consistent with all three
model calculations for either $s$-wave or standard or modified $d$-wave pairing symmetry with large gap $\Delta_0=1.38$ meV.

A line of field-induced quantum criticality in \cerhin\ extends from 1.75 GPa in the zero-magnetic-field limit to 2.3 GPa at the superconducting upper critical field boundary and may influence the Andreev signal.
Since both \cecoin\ and \cerhin\ are inherently multiband heavy-fermion superconductors, additional bands may play a more prominent role due to the fine tuning of Fermi surface nesting of spin fluctuations. While our two-channel tunneling model can account for the Fano resonance, its single conduction band cannot fully capture the intricate interplay between AFM and SC in a truly multiband picture. 
However, as we have shown, within a single-band picture a $d$-wave gap function with a modified slope at the nodes is sufficient to describe the Andreev reflection signal in \cerhin\ at high pressure.
Irrespective of the specific gap function chosen, we find consistently for \cerhin\ a gap amplitude $\Delta_0$ that is more than twice as large as in \cecoin. Thus placing \cerhin\ into the  very strong-coupling regime with $\Delta_0/T_c \sim 6$ compared to $\Delta_0/T_c \sim 2.14$ of a weak-coupling superconductor with $d$-wave gap symmetry.

The effects of quantum criticality on the superconducting pairing correlations are 
often discussed in the context of an antiferromagnetic QCP within the Landau theory of order parameter fluctuations.\cite{Coleman2001,Si2001, Si2003} 
More recently an alternative scenario of local quantum criticality, namely, the critical destruction of the Kondo effect has been proposed to describe $\omega/T$ scaling and a jump in the Fermi surface volume of several heavy fermions, for example, \cerhin.\cite{Pixley2012, Pixley2013, Si2013} 
Such a Kondo-destruction QCP was shown to lead to enhanced superconductivity within a Bose-Fermi Anderson model and therefore might hold the explanation for the drastically enhanced superconducting gap amplitude $\Delta_0$ in \cerhin\ at high pressures compared to \cecoin.

\section{Conclusion}

In summary, we have derived an analytic formula for the point-contact differential conductance of a two-channel tunneling model in the normal and superconducting state. Our generalized two-channel tunneling model has the well-known limits of point contacts both in the normal and superconducting state.
It is applicable to a wide class of materials and not limited to heavy fermions.
In the normal state the two channels of localized surface states and itinerant electrons interfere to create the Fano resonance.
When direct tunneling between the metal tip and 
(heavy-fermion)
superconductor vanishes and instead occurs via the localized state a symmetric Lorentzian line shape is recovered.
In the superconducting state an Andreev reflection signal on top of the asymmetric Fano resonance, which is on top of an additional
large but temperature-independent background conductance, is found 
for model parameters
in the low transparency regime.
Low transparency is expected between materials with large Fermi velocity mismatch, as is the case for tunneling between normal metals and heavy fermions. On the other hand, the low transparency contact does not yield an Andreev reflection signal in the widely used BTK theory.

We have also shown that the SPCS spectra are in quantitative agreement with conventional metal tip PCS data for \cecoin, further validating the application of soft point contacts to heavy-fermion superconductors.
Finally, in \cerhin\ we found that superconductivity is consistent with model calculations of single-band $d_{x^2-y^2}$-wave symmetry, but the opening of the gap at the nodes is drastically modified both near the AFM-SC coexistence region and deeper inside the SC dome. 

The SPCS technique opens up the exciting possibility of studying in detail the evolution of electronic gaps across a quantum critical point with pressure as the control parameter. Because of the intricate interplay between magnetism and superconductivity in the coexistence region,
as well as the superconductivity at high pressures, yet in close proximity to a QCP, a consistent and quantitative analysis of $dI/dV$ curves will
require the inclusion of electron correlation effects into the multiorbital, low-energy model Hamiltonian of heavy-fermion superconductors.

\section{Acknowledgments}
We are grateful to R. Movshovich for stimulating discussions.
M. F.\ was supported by the Swedish Research Council.
Work at the Los Alamos National Laboratory was supported by the US DOE 
under contract No.~DE-AC52-06NA25396 through the LDRD Program (M. J. G.) and
by  the  Office of Basic Energy Sciences, Division of Materials Sciences and Engineering (V. A. S., X. L., E. D. B. and J. D. T.).
M. J. G.\ thanks the Chalmers University of Technology for its hospitality and the Erasmus Mundus Program for travel support.

\section*{Appendix}
In this appendix we briefly construct the tunneling Hamiltonian of a point contact and derive the expressions for current and differential conductance. A detailed derivation can be found in Ref.~\onlinecite{Fogelstrom2010}.
From the outset we wish to emphasize that our model is general and not limited to heavy-fermion systems. Although the importance of localized surface states 
caused by the tip, when pressed into the sample, is most likely only found in correlated electron systems with $d$ and $f$ electrons. On the other hand, 
localized states due to adatoms on the surface or impurities on the tip can be present in any experimental setup either by design or accidentally.

The Eq.~(\ref{eq:tunnelingcurrent}) is derived assuming the following Hamiltonian describing the point-contact setup shown in Fig.~\ref{fig:cartoon_hamiltonian}(a)
\begin{equation}
{\cal{H}}_{\rm tot}={\cal{H}}_{\rm HF}+{\cal{H}}_{\rm tip}+{\cal{H}}_{\rm loc}+{\cal{H}}_{\rm T}+{\cal{H}}_{\rm hyb}
\end{equation}
where
\begin{equation}
{\cal{H}}_{\rm HF}=\sum_{k,\sigma} {\cal E}_{h}(k) c^\dagger_{k\sigma} c_{k\sigma}+\Delta(k)c^\dagger_{k\sigma} c^\dagger_{-k-\sigma}+h.c.
\end{equation}
describes the (heavy-fermion) superconductor including surface scatting processes. We consider the correlated normal state through an effective  band
dispersion ${\cal E}_h(k)$ with renormalized heavy electron masses 
and a possible superconducting state with the order parameter $\Delta(k)$. 
For simplicity, we approximate the HF by a constant density of states 
at the Fermi level. This approximation can be relaxed if needed, but would lead to more complicated expressions for the conductance.
The point-contact material is described by a metallic tip of noninteracting electrons
\begin{equation}
{\cal{H}}_{\rm tip}=\sum_{k,\sigma} {\cal E}_{tip} (k) e^\dagger_{k\sigma} e_{k\sigma}.
\end{equation}
We assume the tip to be a simple metal described by a featureless density of states 
at the Fermi level. To capture the Fano line shape seen in the conductance, we introduce a single localized surface state at $\varepsilon=E_0$, which is described by 
\begin{equation}{\cal{H}}_{\rm loc}=E_0\sum_{\sigma}d^\dagger_\sigma d_\sigma.
\end{equation}
A generalization to many localized states with many different energy levels can also be considered, but does not change the key result of the existence of a Fano or Lorentz resonance.
In this model it is sufficient to treat the single localized state as weakly
coupled to the itinerant electrons in the heavy-fermion conductor via 
\begin{equation}{\cal{H}}_{\rm hyb}=
\sum_{k,\sigma}\lbrack v_{k,\sigma}c^\dagger_{k\sigma}d_{\sigma}
+v^\dagger_{k,\sigma,\sigma^\prime}d^\dagger_{\sigma}c_{k\sigma}\rbrack.
\end{equation} 
Finally,  the tunneling Hamiltonian
\begin{eqnarray}
{\cal{H}}_{\rm T}
&=&\sum_{k,;\sigma}\lbrack t_{k,\sigma}c^\dagger_{k\sigma}e_{k\sigma}
+t^\dagger_{k,\sigma}e^\dagger_{k\sigma}c_{k\sigma}\rbrack
\nonumber\\
&+&\sum_{k,\sigma}\lbrack t_{loc;k,\sigma}d^\dagger_{\sigma}e_{k\sigma}
+t^\dagger_{loc;k,\sigma}e^\dagger_{k\sigma}d_{\sigma}\rbrack 
\end{eqnarray}
describes the tunnelling from the tip into 
either the heavy-fermion material or onto the localized state. We assume for simplicity that momentum and spin are
conserved in a tunnelling event. If needed, this constraint may be relaxed.

The current across the contact can be calculated as 
\begin{equation}
I_{tip/HF}=e\langle {\dot{\hat {\cal{N}}}}_{tip/HF}\rangle =\frac{i e}{\hbar}\langle \lbrack \hat {\cal{H}}_{T},\hat {\cal{N}}_{tip/HF}\rbrack\rangle 
\end{equation}
where ${\hat {\cal{N}}}_{tip/HF}$ is the number operator in the tip or in the heavy-fermion material and $e=-|e|$ is the electron charge. 
It is straightforward to evaluate the commutator for the current into the tip
\begin{eqnarray}
I_{tip}(t,t^\prime)=-\frac{i e}{\hbar} \sum_{k,\sigma} \Bigl\lbrack t_{k,\sigma}\langle e^\dagger_{k\sigma}(t)c_{k\sigma}(t^\prime)\rangle
                                                                  &+&\nonumber\\ 
                                                                  t_{loc;k,\sigma}\langle e^\dagger_{k\sigma}(t)d_{\sigma}(t^\prime)\rangle&-&h.c.\Bigr\rbrack .
\end{eqnarray}
The expectation values define the non-equilibrium Green's functions 
\begin{equation}
G_{\alpha,\beta}^<(k\sigma,k^\prime\sigma^\prime;t,t^\prime)=i\langle c_{\alpha,k\sigma}^\dagger(t) c_{\beta,k^\prime\sigma^\prime}(t^\prime)\rangle,
\end{equation}
where indices $(\alpha,\beta)$ enumerate the different reservoirs, i.e., heavy-fermion or tip material or localized state.
The equal-time current is then cast in terms of the Green's functions
\begin{eqnarray}
I_{tip}(t) &=&
-\frac{e}{\hbar} \sum_{k,\sigma} \Bigl\lbrack \check t_{k,\sigma}\check  G_{tip,hf} (k\sigma;t) +
\nonumber\\ && \qquad
\check t_{loc;k,\sigma}\check G_{tip,loc} (k\sigma;t)-
\nonumber\\ && \qquad
\check t^\dagger_{k,\sigma} \check G_{hf,tip} (k\sigma;t)-
\nonumber\\ && \qquad
\check t^\dagger_{loc;k,\sigma}\check G_{loc,tip} (k\sigma;t) \Bigr\rbrack^K .
\end{eqnarray}
This is the Fourier transform of the current equation (\ref{eq:tunnelingcurrent}) in the main text. 
We include the Nambu-Keldysh-space to include superconductivity, hence 
the checks on the Green's functions and tunnelling matrices.

As the expression for the current involves Green's functions with arguments on either side of the point contact, we need to evaluate these by knowing the Green's functions
in either contact or reservoir or tunnelling matrix elements. This is usually done by writing a formal perturbation theory in the tunnelling elements and summing to infinite order. The summation is performed in the non-crossing approximation, i.e.,  neglecting interference between distinct quasiparticle-tunnelling events
by solving the Dyson equation in reservoir space (the tilde on the Green's function):
\begin{equation}
\check {\tilde G}=\check {\tilde G}_0+\check {\tilde G}_0\circ \check {\tilde T} \circ \check {\tilde G}.
\end{equation}
The $\circ$-product is  short-hand notation for summation or integration over intermediate arguments (energy and momentum) of the Green's functions.
This method is described in detail in text books, see e.g., Refs.~\onlinecite{HaugJauho,Cuevas} and for the current case Ref.~\onlinecite{Fogelstrom2010}. 

It is important to stress that both the Fano line shape of the conductance, Eq.~(\ref{eq:kernel_NS}), and the Andreev signal of the conductance, 
Eq.~(\ref{eq:kernel_SC}), are  results of 
the summation to infinite order in tunnelling processes and cannot be obtained in second order perturbation theory. 
Formulating the charge transport within the non-crossing approximation of the perturbation theory allows us to go from the tunnelling limit (second order) to the open point-contact case (infinite order).
Since point-contact experiments considered in the main text involve several hundereds to thousands of contacts in parallel, we assume that they are uncorrelated and noninteracting.
Finally, the total current or conductance is computed as an average over all possible tunneling channels.

\end{document}